\documentstyle[12pt]{article}
\newcommand{\norm}[1]{{\protect\normalsize{#1}}}
\newcommand{\LAP}
{{\small E}\norm{N}{\large S}{\Large L}{\large A}\norm{P}{\small P}}

\newcommand{\be}{\begin{equation}}
\newcommand{\ee}{\end{equation}}
\newcommand{\bea}{\begin{eqnarray}}
\newcommand{\ena}{\end{eqnarray}}
\newcommand{\beano}{\begin{eqnarray*}}
\newcommand{\enano}{\end{eqnarray*}}
\newcommand{\sect}[1]{\setcounter{equation}{0}\section{#1}}
\newcommand{\vs}[1]{\rule[- #1 mm]{0mm}{#1 mm}}
\newcommand{\hs}[1]{\hspace{#1 mm}}

\newcommand{{\cg}}{\mbox{$\cal{G}$}}
\newcommand{\ch}{\mbox{$\cal{H}$}}
\newcommand{\ck}{\mbox{$\cal{K}$}}

\newcommand{\cn}{\mbox{$\cal{N}$}}
\newcommand{\cu}{\mbox{$\cal{U}$}}

\newcommand{\cw}{\mbox{$\cal{W}$}}
\newcommand{\vph}{\varphi}
\newcommand{\prt}{\partial}
\newcommand{\eps}{\epsilon}

\newcommand{\wh}[1]{\widehat{#1}}
\newcommand{\mb}[1]{\hs{5}\mbox{#1}\hs{5}}
\newcommand{\ie}{{\it i.e. }}
\newcommand{\su}{\mbox{${s\ell(2)}$}}
\newcommand{\sln}{\mbox{${s\ell(n)}$}}

\newcommand{\cub}{\mbox{$\overline{\cal{U}}$}}

\newcommand{\com}{\mbox{$Com (\cg^*_-)$}}
\newcommand{\comv}{\mbox{$Com (\cg_-)$}}
\newcommand{\al}{\alpha}

\newcommand{\cwb}{\mbox{$\overline{\cal{W}}$}}
\newcommand{\scwb}{\mbox{\scriptsize $\overline{\cal{W}}$}}
\newcommand{\moc}{\mbox{$Com_{\scwb} (\cw_-)$}}

\newtheorem{prop}{Property}

\newtheorem{lem}{Lemma}
\newcommand{\prf}{\underline{Proof:}\ }
\newcommand{\cqf}{\\ \rightline{$\Box$}}

\newcommand{\C}{\mbox{\rm\hspace{.042em}\rule{0.042em}{.674em}\hspace{-.642em}
C$\,$}}
\newcommand{\R}{\mbox{\rm\hspace{.042em}\rule{.042em}{.694em}\hspace{-.38em}
R$\,$}}

\newcommand{\NP}[1]{Nucl.\ Phys.\ {\bf #1}}
\newcommand{\PL}[1]{Phys.\ Lett.\ {\bf #1}}

\newcommand{\CMP}[1]{Comm.\ Math.\ Phys.\ {\bf #1}}

\begin{document}
\renewcommand{\thefootnote}{\fnsymbol{footnote}}
\newpage
\pagestyle{empty}
\setcounter{page}{0}


\null
\begin{minipage}{4.9cm}
\begin{center}
{\bf  G{\sc\bf roupe} d'A{\sc\bf nnecy}\\ \ \\
Laboratoire d'Annecy-le-Vieux de Physique des Particules}
\end{center}
\end{minipage}
\hfill
\hfill
\begin{minipage}{4.2cm}
\begin{center}
{\bf G{\sc\bf roupe} de L{\sc\bf yon}\\ \ \\
Ecole Normale Sup\'erieure de Lyon}
\end{center}
\end{minipage}

\begin{center}
\rule{14cm}{.42mm}
\end{center}

\vs{10}

\begin{center}

{\LARGE {\bf Non-polynomial realizations of \cw-algebras}}\\[1cm]

\vs{10}

{\large F. Barbarin$^{1}$, E. Ragoucy$^{1}$, and P. Sorba$^{1,2}$}

{\em Laboratoire de Physique Th\'eorique }\LAP\footnote{URA 14-36
du CNRS, associ\'ee \`a l'Ecole Normale Sup\'erieure de Lyon et \`a
l'Universit\'e de Savoie,

\noindent
$^1$ Groupe d'Annecy: LAPP, Chemin de Bellevue BP 110, F-74941
Annecy-le-Vieux Cedex, France.

\noindent
$^2$ Groupe de Lyon: ENS Lyon, 46 all\'ee d'Italie, F-69364 Lyon
Cedex 07,France.
}\\

\end{center}
\vs{20}

\centerline{ {\bf Abstract}}

\indent

Relaxing first-class constraint conditions in the usual Drinfeld-Sokolov
Hamiltonian reduction leads, after symmetry fixing, to realizations
of \cw\ algebras expressed in terms of all the $J$-current
components.

General results are given for \cg \ a non exceptional simple (finite and
affine) algebra. Such calculations directly provide the commutant,
in the (closure of) \cg \ enveloping algebra, of the nilpotent
subalgebra $\cg_-$, where the subscript refers to the chosen
gradation in \cg. In the affine case, explicit expressions are
presented for the Virasoro, $\cw_3$, and
Bershadsky algebras at the quantum level.

\vfill
\rightline{hep-th/9509088}
\rightline{\LAP-AL-536/95}
\rightline{September 1995}

\newpage
\pagestyle{plain}
\renewcommand{\thefootnote}{\arabic{footnote}}
\setcounter{footnote}{0}

\sect{Introduction}

\indent

In a recent paper \cite{BRS}, the construction of a particular class
of finite $\cw$ algebras has been carried out in a way slightly different from
the usual Hamiltonian reduction (H.R.). The method consists in
following the same steps as the ones considered in the canonical
H.R., except that no more constraints are imposed  on the current
components. Let us be more explicit by reviewing rapidly these
technics.

As widely used these recent years \cite{ORaf}, the determination of a
$\cw$ algebra through H.R. first demands to decompose the simple
Lie algebra $\cg$ under interest into its graded parts\footnote{More
precisely, with respect to
the Cartan generator $H$ of an \su-subalgebra, \cg \ decomposes as:
\[
\cg=\oplus_{i =-h}^h\ \cg_i \mb{with} [H, X_i]=i\ X_i \ \ \ \forall X_i\in\cg_i
\]
We will denote by $\cg_\pm$ the subalgebras  $\cg_+=\oplus_{i>0}\ \cg_i$ and
$\cg_-=\oplus_{i<0}\ \cg_i$.}
relative to an
$s\ell(2)$ embedding in $\cg$:
\be
\cg = \cg_- \oplus \cg_0 \oplus \cg_+
\ee

The general element of $\cg$ will be denoted
\be
J=J^a t_a = J^\al t_\al + J^it_i + J^{\bar{\al}} t_{\bar{\al}}\label{eq:2}
\ee
with the $t_\al$'s and $t_{\bar{\al}}$'s ($\al, \bar{\al}=1$,  ..., dim
$\cg_+ =$ dim $\cg_-$) generating $\cg_+$ and $\cg_-$ respectively,
while the $t_i$'s ($i=1$, ..., dim ${\cg}_0$) form a basis of $\cg_0$.
The $J$-components $J^a$ generate the dual basis in $\cg^*$, on
which can be naturally set a Poisson Kirillov Lie algebra structure
$\{., \ .\}_{PB}$. Then one imposes the $J^\al$ components to take
constant fixed (non zero and zero) real values in such a way that
these constraints constitute a set of (Dirac) first class
constraints. This means that the Poisson bracket of any couple of
them weakly commute, or in other words, the result of their Poisson
commutator vanishes when imposing the constraint conditions. These
first class constraints on $\cg^*_-$ induce a gauge invariance on
the $J^a$'s which can be reformulated on the constrained matrix
\be
J=t_-+J^i t_i + J^{\bar{\al}} t_{\bar{\al}}
\ee
as follows:
\be
J \rightarrow J^g = g \ J g^{-1} + k \prt g . g^{-1}
\label{eq:4}
\ee
with $g$ element of the $G$ subgroup $G_+$, the Lie algebra of which
is $\cg_+$. In the above equation, we have considered the $J^a$'s as
functions of a $z$-variable, $z \in \C$ and $\prt \equiv
\frac{\prt}{\prt z}$. In the case where the $J^a$'s are constant, the gauge
transformation reduces naturally to
a conjugation.

By a suitable gauge fixation, one can transform $J$ into:
\be
J^g=t_-+W^k t_k
\ee
with the $W^k$'s polynomials in the $J^i, J^{\bar{\al}}$ and their
derivatives, and the $t_k$'s constituting a \cg-subset of dim $\cg_0$
generators. The $W^k$'s are gauge invariant;
they weakly P.B. commute with the constraints and form a basis of
the $\cw$ algebra one wishes to construct. Such a $\cw$ algebra will be
a finite $\cw$ algebra \cite{BT} when the $J^a$'s are $z$ independent: in this
last case the $W$ generators can be related to the zero modes of
the $z$-dependent $W^k$ ones.

As emphasized in \cite{BRS}, if one ignores the constraints on $\cg^*$
and acts with the $G_+$ subgroup as in (\ref{eq:4}) on the
unconstrained $J$ given in (\ref{eq:2}), one will then get by a symmetry fixing
$\tilde{W}$ quantities which strongly Poisson commute with the
$J^\al$'s. In \cite{BRS}, we have focussed our attention on finite $\cw$
algebras -as above mentioned, the coadjoint action (\ref{eq:4}) is
then identical to the adjoint one- and also to the case where
$\cg_-$ (resp. $\cg_+$) is Abelian. The so obtained $\tilde{W}$
show up as functions $P (J^a)/Q (J^\al)$ with $P$ a
polynomial in all the $J^a$'s and $Q$ a smooth function in the
$J^\al$'s of $\cg^*_-$ only. Since $\cg_-$ is Abelian and the
$\tilde{W}$ strongly Poisson commute with $\cg^*_-$, the $P
(J^a)$ are in the commutant of $\cg^*_-$. However, one must remark
that the P.B. of two $\tilde{W}$ a priori closes on a
function of the form: $P(\tilde{W}, J^\al)/Q(J^\al)$. That is the
price one has to pay to get a realisation of the $\tilde{W}$
generators in terms of all the $J^a$ components. Let us add that by a direct
quantization, one thus obtains the commutant \comv\ of the subalgebra $\cg_-$
in the closure $\cub(\cg)$ of the \cg-enveloping
algebra $\cu (\cg)$. By closure we mean that we will allow formal series in
the generators of
$\cg_+$, instead of only polynomials. Then, fractions and square roots (for
instance) will be in this
closure.

In addition to the new obtained realization of a $\cw$ algebra, and
the determination of the commutant in ${\cub}({\cg})$ of a nilpotent
subalgebra $\cg_-$, there is another interesting consequence of the relaxing of
the constraints. Indeed, let us suppose we have  a given realization $(R)$
of the algebra $\cg$ in which the $\cg_-$ generators are represented
by -commuting- variables and the other ones in $\cg \setminus \cg_-$ by
differential operators in these variables, or coordinates. Then the
knowledge of the $Com (\cg_-)$ elements, with their explicit
expressions in terms of the $\cg$ generators, allows to deduce from
the canonical representation $(R)$ new realizations of $\cg$. Such
an approach has been exploited in \cite{BRS} on the $Sp (2d, \R)$
algebras in order to reformulate the Heisenberg quantization for a
system of two identical particles in $d=1$ and 2 dimensions,
and to recognize also in $d=2$ a suitable anyonic operator.

\hfill\break

In the present paper, we pursue our investigations and calculations
on any classical simple Lie algebra $\cg$, with a graded
decomposition relative to any of its $s\ell(2)$ embeddings. We widely
develop the finite case, before considering, in the last section, the
affine one with explicit calculations on $\cg = s\ell(2)$ and
$s\ell(3)$.

The first technical problem we are faced with concerns a suitable
symmetry fixing. Indeed the usual Drinfeld-Sokolov highest weight
gauge does not appear adapted in the general case if we do not
impose constraints. We first present in section \ref{Abel} our
symmetry fixing for $\cg = s\ell(n, \R)$ and $\cg_-$ given by the gradation
of the principal
$s\ell(2,\R)$, leaving for an appendix the treatment of an orthogonal
or symplectic Lie algebra $\cg$ submitted to the principal gradation.
Then we generalize the symmetry fixing for any $s\ell(2)$ gradation of
a classical simple Lie algebra (section \ref{sect2}). One can note
that our presentation is facilitated by a graphical matrix description. The
proof is
rather detailed in the case of $\sln$ algebras, and we use the
folding operation allowing to go from the unitary to the
orthogonal and symplectic algebras to study these last ones. Once
again, the obtained $\tilde W$ elements are quotients $P(J^a
) /Q(J^\al)$ with the $Q(J^\al)$ (PB-)commuting with the $\cg_-^*$
elements and also with the $P$'s. By recalling in section
\ref{quant} the direct quantization procedure for finite $\cw$
algebras, we are naturally led to consider the commutants $\comv$ of
the nilpotent subalgebras $\cg_-$.

Actually, the $Q$'s belong to the center of the obtained commutant.
Rather unexpected, the set of all the different $Com
(\cg_-^{\cal H})$, each relative to an $s\ell(2,\R)$ itself principal in a
regular subalgebra $\ch$ of $\cg$, does not provide all the
different $\cw(\cg, \ch)$ algebras. As developed in the section
\ref{sect2.3}, each $Com (\cg_-^{\cal H})$ is isomorphic, up to a center part,
to
the $\cw({\cg}, \mu \ A_1)$ algebra, where $\mu$ is the maximal number
of $A_1$ subalgebras contained in the just defined $\ch$ subalgebra.
Following the type of considered $\cg$ algebra, one has to separate
the cases where the $A_1$'s are all $s\ell(2,\R)$ -$\sln$ type-
from the ones where some of them is an $so(3) $ -$so(n)$
type- or $sp(2,\R)$ -$sp(2n)$ type- algebra. For the reader
surprised by this result -i.e. that all the $\cw(\cg, \ch)$ are not obtained in
this
way-, one may refer to figure \ref{carre} of
section \ref{Abel} where we have tried to graphically represent our
symmetry fixing in the $\cg = s\ell(n)$ case: there, the nested
squares  in the $\cg$ matrix characterize the
commuting $s\ell(2,\R)$ subalgebras, the corresponding matrix elements
of which are lying at the corners of each square.

Now, one may wonder if there is a possibility to get from a given
$\cg$ the other $\cw$ algebras which have not been obtained. This
point is discussed in section \ref{sect6} and illustrated in section
\ref{sl4} by a detailed study of the $s\ell(3)$ and $s\ell(4)$ algebras.
We show on these examples that the not yet obtained $\cw$
algebras can be reached either by enlarging the conjugation
(\ref{eq:4}) to a $\cg$ subalgebra bigger than $\cg_+$, and no more
nilpotent, or by carrying out a secondary reduction process \cite{Wgd,MR}.
This last operation consists in applying the same technics of
symmetry transformation, now with the help of $W$ generators, on an
already obtained $\cw$ algebra.

Our symmetry fixing procedure can directly be extended to the affine
case. This is remarked in section \ref{aff} where, after presenting
the different steps of  all calculation, we give our explicit
realization of the quantum affine $\cw$ algebras relative to
$s\ell(2)$ (Virasoro case) and to $s\ell(3)$ ($\cw_3$ and Bershadsky
ones).

\hfill\break

\sect{Symmetry fixing: the case of the maximal nilpotent subalgebra in
$s\ell(n)$\label{Abel}}

\indent

As emphasized in the introduction, we start with the most general
element\footnote{Stricktly
speaking, we take an $n\times n$ matrix whose
entries on the lower part of the anti-diagonal are non-zero, \ie
$E_{i,n+1-i}\neq0$ $i=1,...,[n/2]$}
of \cg, as given in eq. (\ref{eq:2}). Limiting our study for the moment to
$J^a$'s which do not
depend on any $z$-variable, the coadjoint transformation given in
(\ref{eq:4}) reduces to:
\be
J^g=g\, J\, g^{-1} \mb{with} g\in G_+ \mb{and} LieG_+=\cg_+
\ee

To give a simple idea of the symmetry fixing that we have chosen, we first
consider the case
$\cg=s\ell(n, \R)$ and $H$ its principal gradation. Then, $\cg_+$ is the
subalgebra
generated by all the positive roots of $s\ell(n)$
(maximal nilpotent subalgebra of $s\ell(n)$). Considering the fundamental
representation of $s\ell(n)$,
the elements are $n\times n$ matrices, and
$\cg_+$ is the set of upper triangular matrices (with zeros on the diagonal).

\newpage

\subsection{Direct calculation}

\indent

We propose to prove the following property
\begin{prop}\label{propAb}
Starting with a general element
$J$ of $s\ell(n)$, one can in an unique way determine the parameters of the
transformations associated
to the maximal nilpotent subalgebra $\cg_+$ by requiring $J^g$ (transformed of
$J$) to have zeros in its whole lower triangular part ($\cg_{\leq0}$ part),
except on the
anti-diagonal.
\end{prop}
\prf: We prove this property by a direct calculation, using the principal
gradation $H$ of $s\ell(n)$.
We first consider the root of lowest grade: it is the lowest root
$(-\psi_0)$ of \sln, and its grade is
$(-h)=-(n-1)$.
Then, we define $\cg_+^{(0)}$ as the subalgebra generated by the elements
of $\cg_+$ that
act on $E_{-\psi_0}$:
\be
\cg_+^{(0)}= \{ X\in\cg_+\ s.t.\ [X, E_{-\psi_0}]\neq0\} \label{g+0}
\ee
$\cg_+^{(0)}$ is a subalgebra of the symmetry group $\cg_+$ (see section
\ref{sect1.2}).
As a first step, we show that one can determine the parameters of the
coadjoint transformations
associated to $\cg_+^{(0)}$ by their action on $E_{-\psi_0}$.

\indent

There is a natural decomposition of $\cg_+^{(0)}$ w.r.t. the principal
gradation:
\[
\cg_+^{(0)}= \oplus_{i>0}\ \cg_i^{(0)} \mb{with}
\cg_i^{(0)}=\cg_+^{(0)}\cap\cg_i
\]
Now, if we denote by $\cg_{1-h}^{(0)}=[E_{-\psi_0},\ \cg_1^{(0)}]$, the
definition of
$\cg_+^{(0)}$ insures that dim$\cg_{1-h}^{(0)}$=dim$\cg_1^{(0)}$. Moreover,
the fact
that $(-\psi_0)$ is the root of lowest grade guarantees that the elements
of $\cg_1^{(0)}$ are the only
elements of $\cg_+^{(0)}$ that lead to $\cg_{1-h}^{(0)}$ by coadjoint
action. Thus, the condition
\[
\left.J^g\right|_{\cg_{1-h}^{(0)}}=0
\]
fixes in an unique way the parameters of $\cg_1^{(0)}$ through linear
equations.
This condition still hold
after the action of the other generators of $\cg_+$.

Once $\cg_1^{(0)}$ is fixed, we consider the action of $\cg_2^{(0)}$. Defining
$\cg_{2-h}^{(0)}=[E_{-\psi_0},\ \cg_2^{(0)}]$, it is clear that only the
elements of $\cg_2^{(0)}$
will lead by coadjoint action to $\cg_{2-h}^{(0)}$ (the elements of
$\cg_{1}^{(0)}$
 have not to be considered now). Then, as above, we will fix linearly (and
in an unique way) the parameters
 of   $\cg_2^{(0)}$ through the condition
\[
\left.J^g\right|_{\cg_{2-h}^{(0)}}=0
\]

Recursively, we will fix the parameters of $\cg_i^{(0)}$ through the condition
\be
\left.J^g\right|_{\cg_{i-h}^{(0)}}=0 \mb{where}
\cg_{i-h}^{(0)}=[E_{-\psi_0},\ \cg_i^{(0)}]
\ \ i=1,\dots,n-1
\label{eq.G(i-h)}
\ee
Once the parameters associated to $\cg_j^{(0)}$ $(j<i)$ are fixed, the
condition (\ref{eq.G(i-h)})
 fixes in an unique way (and linearly) the parameters of $\cg_i^{(0)}$.

Thus, we have fixed all the parameters of $\cg_+^{(0)}$ through the condition
 \be
\left.J^g\right|_{\cg_{\leq0}^{(0)}}=0 \mb{where}
\cg_{\leq0}^{(0)}=[E_{-\psi_0},\ \cg_+^{(0)}]
\ee
As we get linear equations, the parameters will be fractions. However, due
to the
definition of $\cg^{(0)}_+$, the denominators will be built on
$J^{-\psi_0}$ only, and we have
supposed that it is not zero.

Considering the
$n\times n$ matrix of \sln, it is not difficult to see that $E_{-\psi_0}$
is in the lower-left corner
of the matrix, while
$\cg_+^{(0)}$ constitute the "edge" of $\cg_+$, \ie elements of the form
$E_{1,j}$ and
$E_{i,n}$. $\cg_{\leq0}^{(0)}$ is almost the edge of $\cg_-$: if
$H_{\psi_0}$ is the Cartan generator
based on $\psi_0$, $\cg_{\leq0}^{(0)}$ is the edge of $\cg_-$ with
$E_{-\psi_0}$ removed, but
$H_{\psi_0}$ included.

\indent

Altogether, if we gather $\cg_+^{(0)}$, $\cg_{\leq0}^{(0)}$ and
$E_{-\psi_0}$ we obtain
the whole border of the $n\times n$ matrix. Then, we are left with a
$(n-2)\times(n-2)$ matrix (that
stands in the middle of the original matrix) and it should be clear that
the second step will deal with
this submatrix. Once again, we determine the root of lowest weight
$(-\psi_1)$ in the submatrix:
by construction of $\cg_+^{(0)}$, $E_{-\psi_0}$ is in the center of
$(\cg_+\setminus\cg_+^{(0)})$, so that at the second step $E_{-\psi_1}$
really plays the role of
$E_{-\psi_0}$ at the
first step. Then we look at  the subalgebra of elements that act on
$E_{-\psi_1}$:
\be
\cg_+^{(1)}= \{ X\in(\cg_+\setminus\cg_+^{(0)})\ s.t.\ [X, E_{-\psi_1}]\neq0\}
\ee
and we will fix the associated parameters through the condition:
 \be
\left.J^g\right|_{\cg_{\leq0}^{(1)}}=0 \mb{where}
\cg_{\leq0}^{(1)}=[E_{-\psi_1},\ \cg_+^{(1)}]
\ee
With the same decomposition w.r.t. the principal gradation ($\cg_+^{(1)}=
\oplus_{i>0}\
\cg_i^{(1)}$ and $\cg_i^{(1)}=\cg_+^{(1)}\cap\cg_i$), we get, grade by
grade, linear equations
that  fix in an unique way the parameters of $\cg_+^{(1)}$. As in the first
step, the denominators of the
fractions appearing in the fields are built only on $J^{-\psi_0}$ and
$(J^{-\psi_1})^g$, where
$(J^{-\psi_1})^g$ is the transformed of $J^{-\psi_1}$ under $G_+^{(0)}$.
Thus, the fields are
well-defined as soon as $J^{-\psi_0}$ and $(J^{-\psi_1})^g$ are not zero.

Once again, we have studied the border of the $(n-2)\times(n-2)$ submatrix,
and we are left with an
$(n-4)\times(n-4)$ submatrix that stands in the middle of the
$(n-2)\times(n-2)$ matrix. Of course,
we continue until we exhaust all $\cg_+$. This ends the proof of the property.
\cqf

\indent

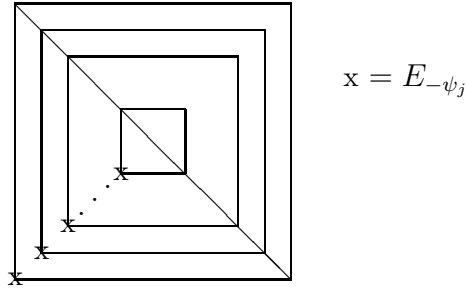
\begin{figure}[hbtp]
\begin{center}
\begin{picture}(164,104)
\put(0,0){\framebox(104,104)}
\put(10,10){\framebox(84,84)}
\put(20,20){\framebox(64,64)}
\put(25,25){\makebox(0,0){.}} \put(30,30){\makebox(0,0){.}}
\put(35,35){\makebox(0,0){.}}
\put(40,40){\framebox(24,24)}
\put(0,0){\makebox(0,0){\small x}} \put(10,10){\makebox(0,0){\small x}}
\put(20,20){\makebox(0,0){\small x}} \put(40,40){\makebox(0,0){\small x}}
\put(0,104){\line(1,-1){104}}
\put(124,75){\makebox(0,0)[l]{{\small x}$\ =E_{-\psi_j}$}}
\end{picture}
\end{center}
\caption{Decomposition in "nested boxes" of the \sln-matrix\label{carre} }
\end{figure}

To visualise the procedure, let us remark that we
have divided the $n\times n$ matrix $J$ into "nested boxes" as drawn in
picture \ref{carre}. In the
lower part (below the diagonal) of the resulting matrix $J^g$ we have only
zeros, except where stand the
different roots $(-\psi_j)$, that is in the lower-left corner of each
"box". In these "corners" and in the
upper part, we have fractions whose denominators are well-defined as soon
as $J^{-\psi_i}\neq0$.

The generalization is quite easy when one realizes that the structure of
"nested boxes" really reflects a
series of mutually embedded subalgebras. To be pedagogical, we first
rephrase the above
calculation.

\subsection{Formal calculation\label{sect1.2}}

\indent

Thus, we come back to the root of lowest grade $(-\psi_0)$. The first thing
to remark is that we have
\be
\begin{array}{l}
\cg_{\leq0}^{(0)}\subset Im(adE_{-\psi_0}) \mb{and} \cg_+^{(0)}\subset
Im(adE_{\psi_0})\\ \\
 Im(adE_{-\psi_0})\cap Im(adE_{\psi_0})= Vect(H_{\psi_0})
 \end{array}
 \ee
where $Vect(H_{\psi_0})$ is the vector space spanned by $H_{\psi_0}$. This
implies that the
"box" formed by $\cg_+^{(0)}$, $\cg_{\leq0}^{(0)}$, and $E_{-\psi_0}$ is
exactly
\be
B_0=Im(adE_{-\psi_0}) +Im(adE_{\psi_0})
\ee
 In fact, we have
 \[
B_0=\cg_+^{(0)}\oplus\cg_{\leq0}^{(0)}\oplus Vect(E_{-\psi_0})
\]
Then, it is a simple exercise to prove that the inside of this box $B_0$ (\ie
the $(n-2)\times(n-2)$ submatrix) is the subalgebra $S_0$ defined by:
\[
S_0=Ker(adE_{\psi_0})\cap Ker(adE_{-\psi_0})
\]
 As vector space, it satisfies:
 \[
 \cg= B_0\oplus S_0
 \]
Moreover, we have the basic property:
\be
\cg_+^{(0)}=\ Im(adE_{\psi_0})\cap Ker(adE_{\psi_0}) \label{cg0}
\ee
which can be taken as a definition for $\cg_+^{(0)}$. Note that this
definition immediately proves
that $\cg_+^{(0)}$ is a subalgebra (since $Im(adX)\cap Ker(adX)$ is always
a subalgebra for any
$X\in\cg$).

Finally, we define recursively the submatrices
at the level $i$ by
\be
S_{-1}=\cg \mb{;}
S_i=\ \bigcap_{j=0}^{i} \left[ Ker(adE_{\psi_j})\cap Ker(adE_{-\psi_j})\right]
\ \ i\geq0
\label{Si}
\ee
Indeed, if we compute the root of lowest grade\footnote{Be careful that the
grading operator has not
changed from the very beginning: it is still the Cartan generator of the
principal \su.} $(-\psi_{i})$,
we will define the symmetry group at the level $i$ by
\be
\cg_+^{(i)}= \left[\ Im(adE_{\psi_i})\cap Ker(adE_{\psi_i})\right] \cap S_{i-1}
\label{cgi}
\ee
and its fixation by
\be
J^g|_{\cg_{\leq0}^{(i)}}=0 \mb{with} \cg_{\leq0}^{(i)}=[ E_{-\psi_i},\
\cg_+^{(i)}] \cap S_{i-1}
\label{Jgi}
\ee
Therefore, after this fixing of the parameters of $\cg_+^{(i)}$, we will
have to work on
\[
S_{i-1} \cap \left[ Ker(adE_{\psi_i})\cap Ker(adE_{-\psi_i})\right] \equiv
S_{i}
\]
while the "box"
\be
B_i= \cg_+^{(i)}\oplus\cg_{\leq0}^{(i)}\oplus Vect(E_{-\psi_i})
\ee
has been completely treated. We have also the property
\be
S_{i-1}= B_{i} \oplus S_{i}
\ee
Although the
definitions (\ref{Si}), (\ref{cgi}) and (\ref{Jgi}) are more formal than
the previously studied, they
insure that at each step the "box" is well-defined and is a subalgebra.
They also prove that
 at each step $i$ we have subalgebras $\cg_+^{(i)}$. Moreover, these
definitions
allow a straightforward generalization as we see in the section \ref{sect2}.

\hfill\break

\sect{Symmetry fixing: the $\cg_-$ general case in a simple Lie
algebra\label{sect2}}

\indent

Let us consider now the general case of a classical simple Lie algebra \cg.
We grade \cg\ w.r.t. a
Cartan generator $H$, itself defined through a \su\ subalgebra. We recall
that, up to few exceptions
(which we discard) and that occur for $\cg=so(2m)$,
this \su\ subalgebra  can be considered as principal in a regular
\cg-subalgebra \ch.
The nilpotent subalgebra $\cn$ whose
 commutant we are looking for, is defined with the help of this gradation
through $\cn=\cg_-$.

The technics to compute the commutant of \cn\ in \cub\
will be the same as in the previous case: we prove that one can find
recursively a
form for $J^g$ that fixes in an unique way and
linearly the parameters of the transformation associated to $\cg_+$.
As the procedure is very similar to the one given in section \ref{sect1.2},
we just mention the
different steps that allow the fixing through some lemmas and properties.

\indent

We start as in the previous case by considering the roots of highest grade
$\psi_p$
(so that $-\psi_p$ is lowest grade). The
difference here is that there can exist more than one such root at each step.
However, generalizing the definition (\ref{cg0}), we
consider the subgroup associated to
\be
\cg_+^{(0)}= \left[ \bigcap_{p=1}^{m} Ker(adE^{(0)}_{\psi_p})\right] \cap
\left[ \biguplus_{p=1}^{m} Im(adE^{(0)}_{\psi_p}) \right]\cap\cg_+
\ee
It is quite easy to see that $\cg_+^{(0)}$ is indeed a subalgebra of $\cg_+$
(using the fact that each $\psi_p$ is
 highest grade).
 One can also prove:
\begin{prop}\label{ceni}
Let \cg\ be a semi-simple Lie algebra, graded w.r.t. a \su-Cartan
generator, and \ch\ the regular
subalgebra defining this \su\ embedding.
We introduce $\Lambda=\{\psi_p,\  p=1,..,m\}$ the set of roots of highest
grade, and
\[
\Lambda_{\|}= \{\psi\in\Lambda \mbox{ such that }E_{-\psi}\in\ch\}\
        =\{\psi_a,\ a=1,\dots,m_{\|}\}
\]
Then, the element
\[
E_{-\vph_0}=\sum_{a=1}^{m_{\|}}E_{-\psi_{a}}
\]
is such that
\be
Ker(adE_{-\vph_0})= \bigcap_{p=1}^{m} Ker(adE_{-\psi_p}) \mb{and}
Im(adE_{-\vph_0})= \biguplus_{p=1}^{m} Im(adE_{-\psi_p}) \label{mino}
\ee
\end{prop}
\prf
We will prove this property for \cg=\sln, and extend it to $\cg=so(m)$ and
$sp(2n)$ by folding.
We start by remarking that we have the following inclusions
\be
\begin{array}{l}
\displaystyle
V_-\subset \left[\bigcap_{p=1}^{m} Ker(adE_{-\psi_p})\right] \subset
Ker(adE_{-\vph_0})\\
\displaystyle
Im(adE_{-\vph_0})\subset \left[\biguplus_{p=1}^{m} Im(adE_{-\psi_p})\right]
\end{array}
\ee
where we have introduced $V_\pm=Vect(E_{\pm\psi_p},\ p=1,..,m)$ the vector
space
(commutative \cg-subalgebra)
generated by the root generators $E_{\pm\psi_p}$.

Moreover, it is easy to see that it is enough to prove one of the
equalities in (\ref{mino}), since we
have:
\be
\begin{array}{l}
\displaystyle
Ker(adE_{-\vph_0})= \bigcap_{p=1}^{m} Ker(adE_{-\psi_p}) \Leftrightarrow
Im(adE_{+\vph_0})= \biguplus_{p=1}^{m} Im(adE_{+\psi_p}) \\
\displaystyle
\Leftrightarrow
Ker(adE_{+\vph_0})= \bigcap_{p=1}^{m} Ker(adE_{+\psi_p}) \Leftrightarrow
Im(adE_{-\vph_0})= \biguplus_{p=1}^{m} Im(adE_{-\psi_p})
\end{array}
\ee
We prove the property by constructing an element $E_{-\vph_0}$ such that
$Im(adE_{-\vph_0})= \biguplus_{p=1}^{m} Im(adE_{-\psi_p})$

\indent

We take \cg=\sln\ and use a graphical method. First, we note that if
$e_{ij}$ is the element $(i,j)$ in
a $n\times n$ matrix, then $Im(ade_{ij})$ is formed by the $i^{th}$ row and
the $j^{th}$ column.
Moreover, if the subalgebra $\ch$ which defines the grading is decomposed as
\be
\ch=\bigoplus_{j=1}^d\ \alpha_j\ s\ell(p_j) \mb{with} p_1>p_2>\dots>p_d>1
\label{eqH}
\ee
there will be more than one root of highest grade if and only if
$\alpha_1\geq2$.
Thus, for the demonstration of the property
\ref{ceni}, we suppose that $\alpha_1\geq2$.

Then, we decompose the set of roots of lowest grade $\Lambda$ into two
orthogonal subsets:
\be
\begin{array}{ll}
\Lambda_{\|} & =\{\psi\in\Lambda \mbox{ such that }E_{-\psi}\in\ch\} \\
 &      =\{\psi_a,\ a=1,\dots,m_{\|}\}
 \end{array}
        \mb{and}
\begin{array}{ll}
\Lambda_{\perp} & =\{\psi\in\Lambda \mbox{ such that }E_{-\psi}\not\in\ch\} \\
 &      =\{\psi_{\bar a},\ {\bar a}=1,\dots,m_{\perp}\}
 \end{array}
\ee
with of course $m_{\|}+m_{\perp}=m$. As a notation, we use $p, q, r,\dots$
for the indices of all
the roots of highest grade, and $a,b,c,\dots$
($\bar{a},\bar{b},\bar{c},\dots$ resp.) to label the roots
that belong to $\Lambda_{\|}$ (resp. $\Lambda_{\perp}$).
We want to show that the generator  $E_{-\vph}=\sum_{a=1}^{m_{\|}} E_{-\psi_a}$
obeys to the property \ref{ceni}.

\begin{figure}[hbtp]
\begin{center}
\begin{picture}(264,164)
\thicklines
\put(0,0){\framebox(164,164)}
\put(0,100){\framebox(64,64)}
\put(64,36){\framebox(64,64)}
\thinlines
\put(54,154){\makebox(0,0){$\bullet$}}
\put(24,124){\makebox(0,0){$s\ell(p_1)$}}
\put(54,10){\line(0,1){144}}
\put(10,154){\line(1,0){144}}
\put(118,90){\makebox(0,0){$\bullet$}}
\put(84,54){\makebox(0,0){$s\ell(p_1)$}}
\put(10,90){\line(1,0){144}}
\put(118,10){\line(0,1){144}}
\put(54,90){\makebox(0,0){$\circ$}}
\put(118,154){\makebox(0,0){$\circ$}}
\put(244,144){\makebox(0,0){$\circ=$ generator of ${\psi_{\bar a}}$}}
\put(244,124){\makebox(0,0){$\bullet=$ generator of ${\psi_{a}}$}}
\end{picture}
\end{center}
\caption{Location of $\psi_a$ and $\psi_{\bar a}$\label{figceni} }
\end{figure}
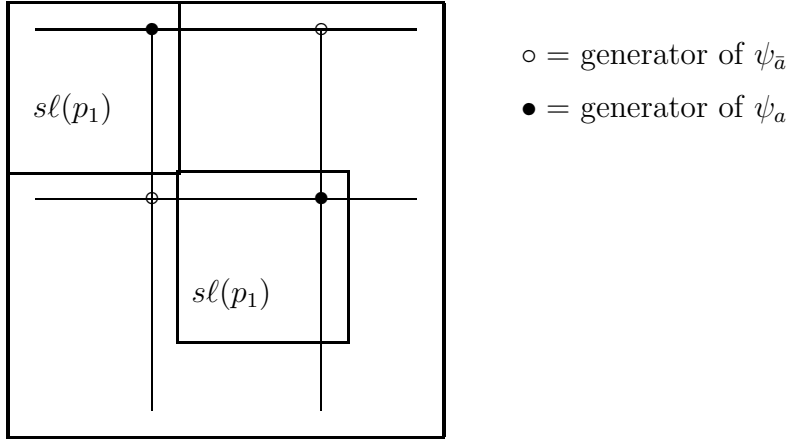

Indeed, it is not difficult to see that any element $E_{-\psi_{\bar a}}$
stands at the crossing of a
row belonging to a set $Im(adE_{-\psi_a})$, with a column belonging to a
set $Im(adE_{-\psi_b})$,
as it is drawn in the figure \ref{figceni}.

Mathematically, this means
\be
\forall\ \psi_{\bar a}\in\Lambda_{\perp}\, ,\ \exists  \{\psi_{a},\
\psi_{b}\}\in\Lambda_{\|}
\mb{such that} E_{-\psi_{\bar a}}\in Im(adE_{-\psi_a})\cap Im(adE_{-\psi_b})
\ee
Since $Im(adE_{-\psi_{\bar a}})$ is formed by two lines (as described
above), this immediately
implies that this last  set is also
part of $Im(adE_{-\psi_a})\uplus Im(adE_{-\psi_b})$, so that
\be
\biguplus_{p=1}^m Im(adE_{-\psi_p})= \biguplus_{a=1}^{m_{\|}} Im(adE_{-\psi_a})
\ee
Now, as the sum in (\ref{eqH}) is direct, it is clear that
\be
Im(adE_{-\vph})= \biguplus_{a=1}^{m_{\|}} Im(adE_{-\psi_a})
\ee
which ends the proof for \sln.

For $so(m)$ and $sp(2m)$, the proof is essentially the same: the only
difference relies on the fact
that root generators have more than one non-zero entries in the
matrices of the fundamental representation. Indeed, we begin with the
matrix realization that we
obtain from the folding of \sln\ matrices (see appendix). The matrices of
$so(m)$ and
$sp(2m)$ are (graded) symmetric w.r.t. the anti-diagonal. Moreover, up to
few exceptional
embedddings that we discard in the case $so(2m)$,
any \su\ embedding in $so(m)$ or $sp(2m)$ can be viewed as the principal
embedding of the regular
subalgebra \ch. This subalgebra itself can always be taken of the form
$\eps\ so(p)\oplus_i
s\ell(p_i)$ for $so(m)$ and $\eps\ sp(p)\oplus_i s\ell(p_i)$ for $sp(2m)$
{\it with $\eps=0$ or 1}.
Then, \ch\
can be considered as resulting from the folding of $\eps\ s\ell(p)\oplus_i
2\ s\ell(p_i)$ in \sln, so that
we can use the graphical method of \sln\ together with the folding.

Thus, for
a root generator $E_{-\psi}$, $Im(adE_{-\psi})$ will be {\it included} in
two lines and two
columns. With a careful identification of the elements of these lines (and
columns) that are not in
$Im(adE_{-\psi})$, one can do the same calculation as above (using the
proof made for
\sln, thanks to the folding property of classical \cw-algebras), so that
the proof for $so(m)$ and
$sp(2n)$ can be carried on.
\cqf

Strictly
speaking, the above calculation is valid for the first step only (with the
same notations as above), \ie
when considering the full \cg\ algebra.
For the other steps $i$, where we deal with
$S_{i-1}=Ker(adE_{-\vph_{i-1}})\cap Ker(adE_{+\vph_{i-1}})$ instead of \cg,
$\ch$ is not a subalgebra of $S_{i-1}$.
However, one has just to consider the restriction $\ch^{(i)}$ of \ch\ to
the \cg-subalgebra $S_{i-1}$, $\ch^{(i)}=\ch\cap S_{i-1}$, as well as the
space $\Lambda^{(i)}$
of roots of lowest grade in $S_{i-1}$.
 Then, decomposing $\ch^{(i)}$ into \su-representations,
$\ch^{(i)}=\oplus_{j=1}^d\ \alpha'_j\ s\ell(p'_j)$, we will get more than one
root of highest grade if and only if $\alpha'_1\geq2$. In that case, we form
the subset of $\Lambda^{(i)}$ associated to
$\ch^{(i)}$:
\[
\Lambda^{(i)}_{\|}= \{\psi^{(i)}\in\Lambda^{(i)} \mbox{ such that
}E_{-\psi^{(i)}}\in\ch^{(i)}\}\
        =\{\psi^{(i)}_a,\ a=1,\dots,m^{(i)}_{\|}\}
\]
and the following generator
\[
E_{-\vph_i}=\sum_{a=1}^{m^{(i)}_{\|}}E_{-\psi^{(i)}_{a}}
\]
will satisfy the property \ref{ceni} at the step $(i)$.

\indent

With the help of property \ref{ceni}, we can now determine a fixing for the
parameters of
$\cg^{(i)}_+$ (where $i$ denotes as above the step we are considering). At
each step, we work on
the subalgebra $S_{i-1}$ of \cg. In this subalgebra, we fix the parameters
of the transformations
associated to $\cg_+^{(i)}$, defined as the $\cg_+$-subalgebra of
generators acting on
$E_{-\vph_i}$:
\be
\cg_+^{(i)}= \left[\ Im(adE_{\vph_i})\cap Ker(adE_{\vph_i})\right] \cap S_{i-1}
\ee
As above, we introduce
 $\cg^{(i)}_{\leq0}=[E_{-\vph_i}, \ \cg^{(i)}_+]$, and we fix the
parameters through
$J^g|_{\cg^{(i)}_{\leq0}} =0$.
It is easy to see that this condition  fixes in an unique way all the
parameters in
$\cg^{(i)}_+$. Moreover, due to the property \ref{ceni}, we can show the
following lemma:
\begin{lem}\label{hop}
Let
$S_i= Ker(adE_{\vph_i}) \cap Ker(adE_{-\vph_i})$ with $i\geq0$, and
$S_{-1}=\cg$.
Then, for any $i\geq0$, we can decompose $S_{i-1}$ as:
\be
S_{i-1}= S_i\oplus B_i \mb{with} B_i=\cg^{(i)}_+ \oplus \cg^{(i)}_{\leq0}
\oplus V^{(i)}_-
\ee
where we have introduced the subalgebras
\[
\begin{array}{l}
\cg_+^{(i)}= \left[\ Im(adE_{\vph_i})\cap Ker(adE_{\vph_i})\right] \cap S_{i-1}
\mb{;} \cg^{(i)}_{\leq0}=[E_{-\vph_i}, \ \cg^{(i)}_+] \\
V_\pm^{(i)}=Vect\{E_{\pm\psi_p^{(i)}}, \ \ p=1,\dots,m\}
\end{array}
\]
\end{lem}
This lemma ensures that we can perform recursively the fixing of the
parameters. In particular, as we
have $V^{(i)}_-\subset B_i$, $\psi_q^{(i+1)}$, the roots of highest grade
in $S_i$ will be
different from the roots $\psi_p^{(i)}$ of $S_{i-1}$. Hence, they will satisfy
$gr(-\psi_p^{(i)})<gr(-\psi_q^{(i+1)})<0$: this guarantees that the
procedure will
end up in a finite number of steps.

Let us summarize the results of sections \ref{Abel} and \ref{sect2} in the
following
property
\begin{prop}
Let \cg=\sln, $so(n)$, or $sp(2n)$, graded w.r.t. the Cartan generator of
a \su\ subalgebra in \cg. Let $\ck=\uplus_i\ \cg_{\leq0}^{(i)}$, where the
subalgebras
$\cg_{\leq0}^{(i)}$ are defined through the property \ref{ceni} and the
lemma \ref{hop}.
Then, imposing $J^g$ to obey the
condition $J^g|_{\ck}=0$  determines in an unique way the parameters of the
symmetry
transformations
\[
J\ \rightarrow\ J^g=gJg^{-1} \mb{where} g\in G_+\mb{with} LieG_+=\cg_+
\]
By this procedure, the non-zero entries appearing in the final form of
$J^g$ have vanishing PB with
the $J^\alpha$'s. They generate the commutant of these elements in the
closure of the enveloping
algebra of the Poisson-Kirillov one generated by the $J‰$'s. With some
abuse of notations, we will
denote this set \com.
\end{prop}

Now, we can formulate two important remarks.

The first one concerns the quantum analogue of the obtained algebra. The
direct quantization technics
that we recall hereafter directly leads to the determination of the
commutant \comv\ of the subalgebra
$\cg_-$.

The second remarks concerns the affine case. The gauge transformation $g$
belonging to $G_+$, so
is the quantity $\prt g g^{-1}$. Since the above symmetry fixing does not
impose any condition on
the $\cg_+$ component of $J^g$, it directly extends to the affine case, by
simple adjonction of the
$\prt g g^{-1}$ term (and restauration of the $z$-dependence).

\newpage

\sect{Quantization \label{quant}}

\indent

Up to now, we have presented a classical version (\ie with Poisson
brackets) of the commutants and
\cw-algebras.
The quantization of the preceeding approach in the case of finite
\cw-algebras has been already
developed in \cite{BRS}. We just recall here the main points and refer to
\cite{BRS} for more
details.

To each generator $J^a$ we associate an operator $\hat{J}^a$, with as usual
the PB replaced by the
commutator. If we stay at the level of the Lie algebra \cg, this
quantization is a Lie algebra
isomorphism. The problem is to extend this isomorphism to the whole
envelopping algebra \cub. In
that case, we have on the one hand a commutative product with an algebraic
structure defined by
Poisson brackets (classical version); and on the other hand a
(non-commutative) operator algebra.
Clearly, as soon as the (classical) product law enters into the game, we
have problems to represent it
at the quantum level. These problems are (partially) cured by replacing
each classical product by a
{\it symmetrized} product at the quantum level. Indeed, with this rule, one
can map the commutant
of a generator of \cg\ (or the commutant of a Lie subalgebra) into its
quantum analogue. This property
is sufficient for our purpose.
Note however that this rule does not avoid the fact that the (vector space)
isomorphism at the level of envelopping algebras is not an algebra
isomorphism anymore.
Therefore, it may happen that structure constants of the quantum commutant
differ from those of the
classical one, even if the realizations (in terms of Lie algebra
generators) are similar.
We gather the technics in the following scheme:
\[
\begin{array}{rclc}
\mbox{Classical} &\rightarrow& \mbox{Quantum}& \mbox{ Isomorphism } \\
J^a &\rightarrow& \hat{J}^a &\\
\textstyle \cg^* &\rightarrow&
\hat{\cg}\equiv\cg & \hat{\cg}\raisebox{-1.2ex}{\
$\stackrel{\textstyle\sim}{\scriptstyle alg.}$\
}\cg^* \\
\{J^a, J^b\} &\rightarrow& {[\hat{J}^a, \hat{J}^b]} &\\
\textstyle \cu  &\rightarrow&
\wh{\cu} & \wh{\cu}\raisebox{-1ex}{\
$\stackrel{\textstyle\sim}{\scriptstyle v.s.}$\ }\cu\\
\textstyle J^{a_1}J^{a_2}\dots J^{a_k} &\rightarrow&
s_k(\hat{J}^{a_1},\hat{J}^{a_2},\dots,\hat{J}^{a_k}) &\\
\textstyle \com &\rightarrow& \comv & \comv
\raisebox{-1ex}{\ $\stackrel{\textstyle\sim}{\scriptstyle v.s.}$\ }\com
\end{array}
\]
where we have indicated in subscripts when the isomorphisms are algebra
($alg.$) or simply vector
space ($v.s.$) isomorphisms. $s_k$ denotes the symmetrized product of $k$
generators. It contains
$k!$ terms and is
normalized in such a way that $s_k(\hat{J}^a,\hat{J}^a,\dots,\hat{J}^a)=
(\hat{J}^a)^k$

Another problem relies on the quotients we form. In the classical approach,
it is possible to define
the inverse of $J^a$, since the product law is commutative. At the quantum
level, it is in general
difficult to deal with the inverse of an operator $\hat{J}^a$. Fortunately,
the inverse we are
considering all belong to $\cg_-$ (see lemma \ref{cier}). Then, the
operators will be constant on each
representation of \com, and thus it is licite to introduce their inverse.

In the following section, we present the generalization of the previous
approach in a classical
framework, but the above remark still applies, so that we have also the
quantization of our finite
\cw-algebras (see also examples in section \ref{sl4}).

\hfill\break

\sect{Commutants and \cw-algebras\label{sect2.3}}
\subsection{Generalities}
\indent

Now that we know how to compute \com\ (and \comv), the problem is to
identify the algebraic
structure of this algebra. In particular, one has to remark that
this algebra  is built on quotients of fields. At the classical level, we
can deal
with such objects, but at the quantum level we have to define fractions of
operators. Fortunately, all
the fractions that appear in our algebra have denominators which are in the
center of the commutant
\com.
\begin{lem}\label{cier}
All the fields $(J^{-\psi_j})^g$ are in the center of \com. These fields
are the only ones
appearing as denominators in the fractions we have considered.
\end{lem}

\prf From the
conditions given in (\ref{Jgi}), it is clear that
the parameters will depend on $\cg_-$ elements only. Moreover, by
construction, $(J^{-\psi_j})^g$
is in the commutant of $\cg_-$ (since it is a non-vanishing matrix element
of $J^g$).
On the other hand, as all the parameters belong to $\cg_-$, and since
$J^{-\psi_j}$ is itself an
element of $\cg_-$, we deduce that $(J^{-\psi_j})^g$ is a function of $\cg_-$
elements only.
Thus, $(J^{-\psi_j})^g$ is in the commutant of $\cg_-$ and formed from
$\cg_-$-elements only: this implies that the set of all $(J^{-\psi_j})^g$
is in the center of
\com.
Since we have already seen in the proof of property \ref{propAb}
that the denominators are functions of the fields $(J^{-\psi_j})^g$ only,
this also proves that these denominators are in the center of \com, and
thus are scalar in each
representation of the algebra.
\cqf

Let us remark that since all the denominators are themselves in \com,
one can choose a finite dimensional basis of \com\ whose elements (say
$P_\alpha$)
 are polynomials in the $J^\alpha$'s and $J^i$'s.
 Using this basis, one can consider the subalgebra of \com\
 generated by polynomials in the $P_\alpha$'s.
 In this subalgebra,
we have only  polynomials in the $J^\alpha$'s and $J^i$'s, and the PB are also
manifestely polynomials in the $P_\alpha$'s. This algebra is then a
polynomial subalgebra not only
of \cub\ but also of \cu, the usual enveloping algebra of \cg.
However,
 one has to be careful that there are polynomials in the $J^\alpha$'s and
$J^i$'s that belongs to
\com\ but that are {\em not} polynomials in the $P_\alpha$'s. In other
words, the \com-basis formed
by the $P_\alpha$'s do not generate polynomially  the commutant of $\cg_-$
in \cu\ (see example in
\cite{nonPol} and also below).

On the other hand, let $Q_\alpha$ be the elements of the (rationnal) basis
one obtains directly
through the fixing of the symmetry. Then, any polynomial in the $J^\alpha$'s
and
$J^i$'s that belongs to \com\ will be expressed as a polynomial in the
$Q_\alpha$'s. Therefore,
the subalgebra formed by the polynomials in the $Q_\alpha$'s will contain
the commutant of
$\cg_-$ in \cu, the usual enveloping algebra of \cg. Thus, we believe that
for the study of the
commutant of $\cg_-$ the $Q_\alpha$'s are more "natural" than the $P_\alpha$'s.

\indent

We have seen that \com\ possesses a center. Apart from this center, we
should identfy the
non-trivial part of \com. For such a purpose, and as for the calculation of
\com, we start with the
simplest case, namely the maximal nilpotent subalgebra in \sln.

\newpage

\subsection{Maximal nilpotent subalgebra in $s\ell(n)$}

\indent

\begin{prop}\label{ouf}
If we decompose \cg=\sln\ w.r.t. its principal gradation, we have the
following algebra
isomorphism:
\be
\com \sim \cw(\sln,\ \mu\ \su)\ \oplus\ \C^\mu \mb{with}
\mu=\left[\frac{n}{2}\right]
\label{eq.guess}
\ee
\end{prop}
Before proving this proposition, let us mention that the center of \com\
contains at least $\mu$
different fields (hence is at least $\C^\mu$), since it contains all the
$(J^{-\psi_j})^g$ fields.
Moreover,
it is clear that \com\ will contains dim\cg$-$dim$\cg_+=
(\frac{n(n+1)}{2}-\mu-1)+\mu$ generators: a direct calculation shows that
there are exactly
$(\frac{n(n+1)}{2}-\mu-1)$ highest weights for the diagonal \su\ in $\mu\
\su\subset_{reg}\sln$.
Therefore,
the dimensions of the two algebras described in (\ref{eq.guess}) are equal.

\indent

\prf: From the final form of $J^g$, it is natural to consider the \su\
subalgebras built on the root
generators $E_{\pm\psi_i}$. These subalgebras mutually commute (due to the
construction of the
$E_{\psi_i}$), so that we get $\mu$ regular $\su$ subalgebras in \sln. To
these $\mu\ \su$, we can
associate a natural gradation, corresponding to the Cartan generator in the
diagonal \su, namely
 $\cg=\cg^{-1}\oplus \cg^0\oplus \cg^{1}$, so that we have a bigrading of
\cg. For instance, under
 the  new  gradation,  the symmetry group $\cg_+$ is divided in
 two  parts $\cg_+= \cg_+^0\oplus \cg_+^+$, where we have indicated in
superscript the new
 grading. In the same way, we get
 $\cg_-= \cg_-^0\oplus \cg_-^-$, and $\cg_0=\cg_0^0$.

 \be
\begin{picture}(107,107)(0,-7)
\put(0,0){\framebox(104,104)}
 \put(75,75){\makebox(0,0){$\cg_+$}} \put(30,30){\makebox(0,0){$\cg_-$}}
\put(0,104){\line(1,-1){104}}
\put(116,-14){\vector(-1,1){10}}\put(116,-14){\line(1,0){8}}
\put(130,-8){\makebox(0,0){$\cg_0$}}
\end{picture}
\begin{picture}(64,107)(0,-7)
\put(24,54){\vector(1,0){24}}
\end{picture}
\begin{picture}(107,107)(0,-7)
\put(0,0){\framebox(104,104)}
\put(84,42){\makebox(0,0){$\cg_+^0$}} \put(42,84){\makebox(0,0){$\cg_+^0$}}
\put(75,75){\makebox(0,0){$\cg_+^+$}}
\put(20,62){\makebox(0,0){$\cg_-^0$}} \put(70,18){\makebox(0,0){$\cg_-^0$}}
\put(30,30){\makebox(0,0){$\cg_-^-$}}
\put(52,0){\line(0,1){104}} \put(0,52){\line(1,0){104}}
\put(0,104){\line(1,-1){104}}
\put(116,-14){\vector(-1,1){10}}\put(116,-14){\line(1,0){8}}
\put(130,-8){\makebox(0,0){$\cg_0^0$}}
\end{picture} \label{fig2}
\ee

 Using this decomposition, we make the symmetry group $G_+$ acting in two
steps.

 In the first step, we consider $\cg_+^0$.  Using the bigrading, it is easy
to see that it is a
 subalgebra. We use the associated subgroup to set $J$ into the form:

\be
J'=\left(\
\begin{picture}(107,60)(0,50)
\put(0,0){\framebox(104,104)}
\put(0,104){\line(1,-1){104}}
\put(0,52){\line(1,0){52}} \put(52,0){\line(0,1){52}}
\put(0,0){\makebox(0,0){*}} \put(10,10){\makebox(0,0){*}}
\put(25,25){\makebox(0,0){.}} \put(30,30){\makebox(0,0){.}}
\put(20,20){\makebox(0,0){.}}    \put(40,40){\makebox(0,0){*}}
 \put(52,52){\makebox(0,0){*}}
\put(10,40){\makebox(0,0){0}}\put(40,10){\makebox(0,0){0}}
\end{picture} \right)
\label{truc}
\ee
Let us first remark that the bigrading implies
$[\cg_+^0,\cg_-^-]\subset\cg_-^-$. Moreover,
$\cg_+^0$ has dimension $\mu(\mu-1)$ while $dim\cg_-^-=\mu^2$. As the
anti-diagonal in
$\cg_-^-$ is not fixed, we need $\mu^2-\mu$ group generators to get the
form (\ref{truc}), and it is
easy to convince one-self by a direct calculation that $\cg_+^0$ indeed
acts correctly on $\cg_-^-$ for
that purpose.

Now, in the second step, we will deal with the subgroup $\cg_+^+$. This
subgroup is exactly the
symmetry group associated to the reduction w.r.t. $\mu\ \su$, and the
matrix (\ref{truc}) has just the
form of the corresponding constrained matrix (up to the grade $-1$ terms
$J_{-\psi_i}$ which have
been left free). This kind of reduction (with an Abelian symmetry group)
has been studied in
\cite{BRS}, and we know that the resulting algebra is just the
$\cw(\sln,\mu\ \su)$ algebra
plus a center that
corresponds to the $J_{-\psi_i}$'s themselves. This ends the proof.
\cqf

 Let us remark that the decomposition $\cg_+= \cg_+^0\oplus \cg_+^+$
divides $\cg_+$ in a part
 ($\cg_+^0$) that acts on the $J_{-\psi_i}$'s, and an abelian part
($\cg_+^+$) that does not
 transform  them.

\subsection{The other nilpotent subalgebras in \sln}

\indent

The above procedure can be applied to any $\cg_-$, subalgebra of \cg\
relative to a given
\su-embedding.
We introduce a new gradation (written in superscript) for \cg:
$\cg=\cg^{-1}\oplus \cg^0\oplus
\cg^{+1}$ that is directly related to the \cw-algebra we will recognize
inside the commutant. Thus,
we have a bi-gradation of \cg\  (cf figure \ref{fig2})
and the property:
\begin{prop}
Let $\rho$ be the total number of roots $\psi_p^{(i)}$ ($\forall p,\forall
i$), and $\mu\leq\rho$ the
number of roots $\psi_a^{(i)}$ that enter in the decomposition of $\vph_i$'s:
\[
\rho=\sum_i\ m^{(i)}\mb{and} \mu=\sum_i\ m^{(i)}_{\|}
\]
Then, we have the following isomorphism
\[
 \com \sim \cw(\sln,\ \mu\ \su)\ \oplus\ \C^\nu  \mb{with}
 \mu\leq\nu= 2dim\cg_+^+ - dim\cg_+\leq\rho
\]
and $\mu$ will be also the maximal number of regular \su\ that one can
simultaneously embed in \ch, the algebra that defines the gradation.
\end{prop}

\prf We have first to identify the \su\ embeddings we consider. Since at
each step of the process we
may have several highest grade roots, we use the property \ref{ceni} to
distinguish one from the
others,  \ie
we select the $E_{\pm\vph_i}$'s that satisfy the equation (\ref{mino}).
Then, it is clear that the \su\
algebras built on these roots will commute, since $B_i$ is in direct sum
with $B_j$ in \cg\ as soon as
$i\neq j$. Now, we have to calculate the number of {\it regular} \su\
subalgebras
we get. It is not simply the number of $E_{\vph_i}$ generators because
these elements may be not
regular in \cg. Thus, one has to consider the regular generators
$E_{\psi_p}$. Therefore,
at each step, we have
 just the number of root generators that enter in the decomposition of
$E_{\vph_i}$ (since these roots mutually commute). Thus, at each step we
get $m^{(i)}_{\|}$ \su\
subalgebras (same notation as in property \ref{ceni}), and the total number is
$m_{\|}=\sum_i\ m^{(i)}_{\|}$.

 Now, using the gradation associated to the diagonal \su, one remarks that each
$\cg_+^{(i)}$ divides into
$(\cg_+^{(i)}\cap\cg^0)\oplus(\cg_+^{(i)}\cap\cg^+)$ and that the
associated subgroup does not interact in the determination of the
parameters (since the lowest grade
roots have a grade -1 w.r.t. this new gradation).  Therefore, we
can make $(\cg_+\cap\cg^0)$ act first, and use it to set $J$ into the form
\be
J\ \rightarrow\ J_{-1}+J_{\geq0}
\ee
where the index denotes the grade.

Then, the
group associated to $\cg^+_+=(\cg_+\cap\cg^+)$, which is Abelian, is simply
the group
corresponding to the reduction w.r.t. $\mu\ \su$ and we end as in property
\ref{ouf}.

The dimension of the center subset $\C^\nu$ is then determined by a
counting argument. The
symmetry group has dimension $dim\cg_+$, so that \com\ has dimension
$dim\cg-dim\cg_+$.
The algebra
$\cw(\cg , \mu\ \su)$ has dimension $dim\cg-2dim\cg^+_+$, since it can be
obtained through
Hamiltonian reduction with gauge group $\cg^+_+$. Therefore, this center
part has dimension
$(dim\cg-dim\cg_+)-(dim\cg-2dim\cg^+_+)=2dim\cg^+_+-dim\cg_+$.
\cqf

Let us remark that
in the case of the maximal nilpotent subalgebra, we have $\mu=\nu=\rho$, the
number of
roots $\psi_i$ since there is only one lowest grade root at each step.

We end this paragraph by emphasizing
  that the \cw-algebra we obtain from the commutant \com\ by disgarding (a
part of) its center
is entirely determined by $\cg_-^-$, the maximal abelian subalgebra of $\cg_-$.

\subsection{Commutants in $sp(2n)$ and $so(m)$}

\indent

The study of commutants in $sp(2n)$ can be immediately deduced from the
\sln\ case. Indeed, using
the folding of $s\ell(2n)$, one realizes that all the technics developed
for \sln\ also applies to
$sp(2n)$.

The calculation for $so(m)$ algebras is slightly different. Indeed, if we
consider the folding of
$s\ell(2n+1)$ to get $so(2n+1)$, we obtain a matrix representation with
zeros on the anti-diagonal.
Then, it is clear that the fixed form for $s\ell(2n+1)$ will not be correct
for the $so(2n+1)$. What
one has to do is to first fold the non-fixed matrix, and then compute the
form one has to take (see
appendix).

Altogether, we have:
\begin{prop}
Let $\cg$ be $so(n)$ or $sp(2n)$, graded with respect to some
$s\ell(2)$-Cartan generator. Let $\rho$ be the total number of roots
$\psi_p^{(i)}$ ($\forall p,\forall i$), and $\mu\leq\nu$ the
number of roots $\psi_a^{(i)}$ that enter in the decomposition of $\vph_i$'s:
\[
\rho=\sum_i\ m^{(i)}\mb{and} \mu=\sum_i\ m^{(i)}_{\|}
\]
Then, we have the following isomorphisms
\[
\begin{array}{l}
\mbox{For} \ \cg = sp(2n):\ \ \ \ \com \sim \cw(sp(2n),\ \mu_1\
\su\oplus\mu_2\ sp(2))\ \oplus\ \C^{\nu}\\
 \mbox{For} \ \cg = so(n):\ \ \ \ \com \sim \cw(so(n),\ \mu_1\
 \su\oplus\mu_2\ so(3))\ \oplus\ \C^{\nu}
\end{array}
 \]
 with $\mu_2=0 \mbox{ or }1$ ; $\mu=\mu_1+\mu_2\leq\nu\leq\rho$
 and $\mu$ will be also the maximal number of regular \su\ and $sp(2)$
(resp. $so(3)$)
that one can
simultaneously embedd in \ch, the algebra that defines the gradation.
\end{prop}
Note that in the property, we have used the fact that in $sp(2m)$ ($so(m)$
respectively), the
regular subalgebras that classify
the \su\ embeddings are sums of \sln's and eventually {\it one} $sp(2)$
(respectively $so(3)$)
subalgebra.

\subsection{Casimirs of \cw-algebras \label{Cas}}

\indent

We propose to show that $det(J^g - \lambda I)$ with $J^g=g J g^{-1}=W^k t_k$,
provides
directly the Casimirs of the \cw-algebra. For such a purpose,
let us  calculate $det(J^g - \lambda I)$ in two different ways.

On the one hand, anticipating on the results of section \ref{norm},
where the elimination of the $\C^{\nu}$ center is considered,
$J^g$ is a matrix containing, apart from $0$ and $1$-entries,
the $W$ generators. Therefore, $det(J^g - \lambda I)$ is a polynomial in
$\lambda$
whose coefficients are expressed as polynomials in the $W$ generators.

On the other hand, $det(J^g - \lambda I) = det(g J g^{-1} - \lambda I)
= det(g (J - \lambda I)g^{-1}) = det(J - \lambda I)$. This is a
polynomial in $\lambda$ whose coefficients commute with all the generators
of the Lie algebra. In the $\sln$ case, $det(J - \lambda I)$ can be
written in the form
$(-1)^n \lambda ^n + \sum_{i=0}^{n-2} C_{n-i} \lambda ^i$ where
$C_i, i=2,\dots,n$ are the Casimirs of $\sln$ (note that the coefficient
of $\lambda ^{n-1}$ is zero since $Tr(J)=0$).

Thus, we have constructed some polynomials in the $W$ generators that can be
expressed as the Casimirs of the Lie algebra. Since the Casimirs commute
with all the generators of the Lie algebra and since the $W$ generators are
all expressed in terms of the Lie algebra generators, the polynomials we
have constructed commute finally with all the generators of the $\cw$-algebra.

We remark that the above property can be used to replace some of the $W$
generators by the
Casimirs by inverting (when it is possible) the expression of the $C_i$'s
in terms of the $W$'s.
Then, the $C_i$'s decouple and participate to the center (of the
\cw-algebra), which thus appears
naturally.

\hfill\break

\sect{Commutant of non-nilpotent subalgebras \label{sect6}}

\indent

Up to now, we have considered the case where \cn\ is nilpotent. We have
obtained in this way many
but not all of the known \cw-algebras. We want to show that the other
\cw-algebras can be also
obtained as commutant of some \cg-subalgebra if one relaxes the condition
of nilpotency, or in other
words take \cn\ as a subalgebra of the full \cg-algebra, instead of $\cg_+$
only. Before coming to the
general case, we show that relaxing the condition of nilpotency already
helps a lot for the calculation
of the \cw-algebras we have obtained.

\subsection{Use of the Cartan subalgebra \label{norm}}

\indent

Let $\cg^-_-$ be an abelian subalgebra, and $\cg_-$ the maximal
nilpotent $\cg$-subalgebra whose maximal abelian subalgebra is $\cg^-_-$.
Then, even in this case where the $\cg_- \setminus \cg^-_-$-part is the biggest
one can get,
$Com_g(\cg_-)$ still contains some central elements, in addition
to those of $\cw(\sln,\mu \su)$. To get rid of this center part, the idea
is to extent the symmetry transformation $\cg_+$ with some Cartan
generators.

Actually, the elements of this center are polynomials, formed with $\cg_-$
generators, \ie only with negative roots. So there exist Cartan generators
such that their eigenvalues under these polynomials are not all zero.
One can choose suitably a set $\tilde{\ch}$ of (2 dim$\cg_-^-$ $-$
dim$\cg_-$) such elements and
use them as symmetry generators. As a result, one will get the commutant in
\cub\ of
$(\cg_-\oplus \tilde{\ch})$, and the undesired center will drop out. One
must note that the current
components which undergo such Cartan transformations can be fixed to $1$ but
not to $0$.

Note that there may be different sets $\tilde{\ch}$; to each choice of
$\tilde{\ch}$ will
correspond a different normalization of the remaining $W$ generators.
Such a phenomenon will be considered in the $s \ell (3)$ case for
the algebra noted $\cw_3^{(2)}$ (see section \ref{W32}).

\subsection{The general case \label{Tgle}}

\indent

In the previous construction of $\cw(\sln,\mu \su)$, apart from
the subalgebra $\cg^+_+$, we act with some other elements, belonging to
$\cg^0_+$.
In fact, $\cg^0_+$ can
be used to transform the $J_{-1}$-part into a constrained
form, identical to the one used in the Hamiltonian reduction formalism.
Consequently, in order to obtain  $\cw(\cg,\ch)$-algebras, with
$\ch\neq\mu\ \su$,
 the idea is to act on $J$ first with elements which
do not belong to $\cg_+$ in order to recover the constrained form for the
$J_-$-part,
and then with $\cg_+$, using for instance the highest weight gauge.
The $\cw$-algebra is then viewed as the
commutant in $\cub$ of a subalgebra $\tilde{ \cg}$ of $\cg$, formed by
$\cg_-$ and
a subset of $\cg_{\geq 0}$, noted $\tilde{ \cg}_{\geq 0}$.

Let  us be more precise. If $J_{const}$ denotes a set of constraints that
leads to \cw(\cg, \ch) in the
usual Hamiltonian reduction, we will determine the $\tilde{\cg}_{\geq 0}$
subset by requiring
\be
J^g|_{<0} = J_{const} \mb{with} g\in \tilde{G}_{\leq 0}
\ee
where $J^g$ is the transformed current under symmetry transformations
belonging to
$(\tilde{ \cg}_{\geq 0})^*=\tilde{ \cg}_{\leq 0}$. Moreover, $\tilde{
\cg}_{\geq 0}$ must also be
such that $\tilde{ \cg}_{\geq 0}\oplus\cg_-$ is a \cg-subalgebra.

\indent

As an example, we consider the \cw(\sln, \sln) case. The $\cg_-$ part
consists of elements below the
main diagonal of the $n\times n$ matrix. The $\tilde{ \cg}_{\geq 0}$ part
can be made with all the
other generators except the ones expressed by the matrices $E_{i,n}$ with
$i=1,\dots, n-1$.
The $\tilde{G}_{\leq 0}$ transformation parameters will be determined
through the condition
\be
(J^g)_{ij} = \delta_{i,j+1} \ \ \  \forall\ j<i \mb{with} g\in
\tilde{G}_{\leq 0}
\ee


\subsection{Commutants of second order \label{watts}}

\indent

Since $\cw$-algebras can be viewed as deformations of semi-simple Lie
algebras, we
may think of performing on those objects the same kind of transformations
we use on Lie algebras to obtain $\cw$-algebras. In the framework of usual
Hamiltonian reductions,
this leads to the so-called secondary reduction \cite{Wgd,MR}.
Here in the same way as in the previous sections, we replace the
Hamiltonian reduction by the
computation of a commutant. More precisely, considering the closure \cwb\
of a \cw-algebra, we
compute the commutant of a given nilpotent  (not necessarily Lie-)
\cw-subalgebra $\cw_-$: in the
following, we will denote this commutant \moc.

\indent
Examples of such secondary paths will be given in section \ref{sl4} for the
$s \ell (3)$ and $s \ell (4)$
cases.

\section{Commutants and \cw-algebras in $s \ell (3)$ and $s \ell (4)$
\label{sl4}}
We deal with the $s \ell (3)$ and $s \ell (4)$ cases in a detailed way.
The $s \ell (3)$ case is the most well-known and the easiest one to
handle with. Although calculations are rather heavy, we decide to
present the $s \ell (4)$ case because it seems to be a good example
providing computation of commutants of nilpotent and non-nilpotent subalgebras
and commutants of second order.
At the classical (quantum) level, generators are denoted by $J$
(respectively $\hat{J}$). In what follows, we will present the quantum
version of each \cw-algebra, since the classical one can be deduced from
the former one by eliminating terms corresponding to quantum corrections;
these last ones can be recognized easily: they have lower degree than the
other terms. In
the quantum framework, we will always consider the symmetrized product
$s_k( \, , \, , )$.

\subsection{$s \ell (3)$ case \label{sl3}}
Let $\alpha,\beta$ be the simple positive roots.
The most general element of $s \ell (3)$ can be written as:
\[
J = \left(
\begin{array}{ccc}
J_1+\frac{J_2}{2} & J_{\alpha} & J_{\alpha + \beta} \\
J_{-\alpha} & -J_2 & J_{\beta} \\
J_{-\alpha-\beta} & J_{-\beta} & -J_1+\frac{J_2}{2} \\
\end{array}
\right).
\]
In $s \ell (3)$, there are two types of regular subalgebras:
$\ch = \su, s \ell (3)$, and so two different \su-embeddings, which
induce two different gradations of $s \ell (3)$,
$\cg=\cg_-^{{\cal H}} \oplus \cg_0^{{\cal H}} \oplus \cg_+^{{\cal H}}$. In the
usual
Hamiltonian reductions, one gets the following \cw-algebras:
\[
\cw_3^{(1)} = \cw(s \ell (3),s \ell (3)) \qquad
\cw_3^{(2)} = \cw(s \ell (3),s \ell (2)).
\]
We present explicitely the computation of
$Com(\cg_-^{{\cal H}})$, for $\ch = \su, s \ell (3)$, and show
that the resulting algebra is $\cw_3^{(2)}$ up to a center, as expected
from section \ref{sect2}. We also display a normalization
procedure in order to get rid of this center. Then we characterize the
$\cw_3^{(1)}$ algebra as a commutant of an $s \ell (3)$ subalgebra as well as
the commutant of a $\cw_3^{(2)}$ subalgebra.

\subsubsection{$\ch=s \ell (3)$ \label{W32}}
In this case $\cg^*_-=\{ J_{-\alpha},J_{-\beta},J_{-\alpha-\beta} \}$.
We consider a coadjoint transformation $J^g=gJg^{-1}$ under a $G^+$-element:
\[
g = \left(
\begin{array}{ccc}
1 & A & C \\
0 & 1 & B \\
0 & 0 & 1 \\
\end{array}
\right).
\]
The generator $J_{-\alpha-\beta}$ associated to the root of lowest grade
with respect to this
principal gradation is invariant under $G^+$ transformations. The
conditions
\[
J^g_{2,1}=0 \, , \, J^g_{3,2}=0 \, , \, J^g_{1,1}=J^g_{3,3}
\]
uniquely fix the parameters $A,B,C$ through linear equations and $J^g$
has the following form:
\[
J^g = \left(
\begin{array}{ccc}
W_0 & W_a & W_{ab} \\
0 & -2W_0 & W_b \\
J_{-\alpha-\beta} & 0 & W_0 \\
\end{array}
\right).
\]
$J_{-\alpha-\beta}$ commutes with all the generators, so the commutant writes
\[
Com(J_{-\alpha},J_{-\beta},J_{-\alpha-\beta}) =
\{W_0,W_a,W_b,W_{ab} \} \oplus \{ J_{-\alpha-\beta} \}
\]
where denominators of $W_0,W_a,W_b,W_{ab}$ only contain the element
$J_{-\alpha-\beta}$.
\be
W_0=\frac{P_0}{2 J_{-\alpha-\beta}} \qquad
W_a=\frac{P_a}{2 J_{-\alpha-\beta}^2} \qquad
W_b=\frac{P_b}{2 J_{-\alpha-\beta}^2} \qquad
W_{ab}=\frac{P_{ab}}{4 J_{-\alpha-\beta}^3} \label{7.1}
\ee
Using the quantization procedure recalled in section \ref{quant}
(\ie symmetrisation) we get the finite quantum \cw-algebra generated by:
\beano
\hat{P}_0 &=& s_2(\hat{J}_2,\hat{J}_{-\alpha-\beta}) +
s_2(\hat{J}_{-\alpha},\hat{J}_{-\beta})\\
\hat{P}_a &=& -2
{\,}s_3(\hat{J}_1,\hat{J}_{-\beta},\hat{J}_{-\alpha-\beta}) -3 s_3(\hat{J}_2,
\hat{J}_{-\beta},\hat{J}_{-\alpha-\beta}) -2
s_3(\hat{J}_{-\alpha},\hat{J}_{-\beta},\hat{J}_{-\beta})
+2 s_3(\hat{J}_{\alpha},\hat{J}_{-\alpha-\beta},\hat{J}_{-\alpha-\beta})
\\
\hat{P}_b &=& 2
s_3(\hat{J}_1,\hat{J}_{-\alpha},\hat{J}_{-\alpha-\beta}) -3 s_3(\hat{J}_2,
\hat{J}_{-\alpha},\hat{J}_{-\alpha-\beta}) -2
s_3(\hat{J}_{-\alpha},\hat{J}_{-\alpha},\hat{J}_{-\beta})
+2s_3(\hat{J}_{\beta},\hat{J}_{-\alpha-\beta},\hat{J}_{-\alpha-\beta})\\
\hat{P}_{ab} &=& 4
s_4(\hat{J}_1,\hat{J}_1,\hat{J}_{-\alpha-\beta},\hat{J}_{-\alpha-\beta})
-6 s_4(\hat{J}_2,\hat{J}_{-\beta},\hat{J}_{-\alpha},\hat{J}_{-\alpha-\beta})
-3 s_4(\hat{J}_{-\beta},\hat{J}_{-\beta},\hat{J}_{-\alpha},\hat{J}_{-\alpha})+
\\
&& +4
s_4(\hat{J}_{-\alpha},\hat{J}_{\alpha},\hat{J}_{-\alpha-\beta},\hat{J}_{-\alpha-\beta})
+4 s_4(\hat{J}_{\alpha+\beta},\hat{J}_{-\alpha-\beta},\hat{J}_{-\alpha-\beta},
\hat{J}_{-\alpha-\beta})+\\
&& +4
s_4(\hat{J}_{-\beta},\hat{J}_{\beta},\hat{J}_{-\alpha-\beta},\hat{J}_{-\alpha-\beta})
\enano

By a simple change of basis, $\hat{W}_{ab}$ can be
replaced by the central element $\hat{C}_2$,
\[
\hat{C}_2 = 3 \hat{W}_0^2 + \hat{J}_{-\alpha-\beta} \hat{W}_{ab}
+\frac{53}{48},
\]
using determinant equalities\footnote{Beware: the determinant property works at
the classical level;
we quantize afterwards.}
(see section \ref{Cas}). Actually, $\hat{C}_2$ is second degree
$s \ell (3)$ Casimir (its explicit expression will be given below).
We note that the constant term $\frac{53}{48}$ comes from quantum corrections.
 In the basis of polynomials
$\{\hat{P}_0,\hat{P}_a,\hat{P}_b,\hat{C}_2,\hat{J}_{-\alpha-\beta} \}$,
commutation relations take the form:
\[
[\hat{P}_0,\hat{P}_a]=\hat{J}_{-\alpha-\beta} \hat{P}_a \qquad
[\hat{P}_0,\hat{P}_b]=-\hat{J}_{-\alpha-\beta} \hat{P}_b
\]
\[
[\hat{P}_a,\hat{P}_b]=-12 \hat{J}_{-\alpha-\beta} \hat{P}_0^2
+\hat{J}_{-\alpha-\beta}^3 (4 \hat{C}_2+3).
\]
Normalization (see section \ref{norm}) consists in eliminating
$\hat{J}_{-\alpha-\beta}$ in those
equations. Since it commutes with all the other quantities, we
can redefine without ambiguities
\[
\hat{Y} =
\frac{\hat{P}_0}{\hat{J}_{-\alpha-\beta}} \qquad
\hat{W}_+ = \frac{-\hat{P}_a}{2 \hat{J}_{-\alpha-\beta}^{3-m}} \qquad
\hat{W}_- = \frac{\hat{P}_b}{2 \hat{J}_{-\alpha-\beta}^{m}} \mb{with}
m \in \R.
\]
In this form we then recognize the commutation relations of the algebra
$\cw_3^{(2)}$, seen as a deformed $\su$ algebra plus a central $\hat{C}_2$
element:
\be
[\hat{Y},\hat{W}_{\pm}] = \pm \hat{W}_{\pm} \qquad [\hat{W}_+,\hat{W}_-]=3
\hat{Y}^2 -(\hat{C}_2+\frac{3}{4}) \label{7.2}
\ee
The analogue of the Casimir for this $\cw$-algebra can be found thanks
to the determinant property and is indeed equal in this realization to
$\hat{C}_3$ the third order Casimir of $s \ell (3)$:
\[
\hat{C}_3 = s_2(\hat{W}_+,\hat{W}_-) +\hat{Y}^3 -\hat{Y}
(\hat{C}_2+\frac{1}{2}).
\]
When $m$ takes the value $\frac{3}{2}$, the quantities
$\hat{Y},\hat{W}_{\pm},\hat{C}_2$ commute
with $\hat{J}_1$. Actually, they generate
$Com(\hat{J}_{-\alpha},\hat{J}_{-\beta},\hat{J}_{-\alpha-\beta},\hat{J}_1)$
that can be obtained by action of a coadjoint transformation
\[
g = \left(
\begin{array}{ccc}
D & A & C \\
0 & 1 & B \\
0 & 0 & \frac{1}{D} \\
\end{array}
\right)
\, \, \mbox{and imposing} \, \, \,
J^g = \left(
\begin{array}{ccc}
Y & W_+ & C_2+Y^2 \\
0 & -2 Y & W_- \\
1 & 0 & Y \\
\end{array}
\right).
\]
 In this case, the degree of the four generators $Y,W_{\pm},C_2$ are
respectively
$1,\frac{3}{2},\frac{3}{2},2$, and corresponds exactly to the spin of the
generators
of the Bershadsky algebra in the affine case.

\subsubsection{$\ch = \su$}
In this case, after halving, $\cg_-=\{ J_{-\alpha},J_{-\alpha-\beta} \}$
is abelian and has gradation $-1$. We will act on $J$ with coadjoint
transformation of the form
\[
g = \left(
\begin{array}{ccc}
1 & A' & B' \\
0 & 1 & 0 \\
0 & 0 & 1 \\
\end{array}
\right).
\]
Now we have more than one root of lowest weight.
Following notations of section \ref{sect2}, we have
$\Lambda_{\Vert}=\{\alpha+\beta \}$.
The symmetry fixing
is determined by $J^g_{3,2}=0 \, , \, J^g_{1,1}=J^g_{3,3}$:
\[
J^g = \left(
\begin{array}{ccc}
W_0^{'} & W_a^{'} & W_{ab}^{'} \\
J_{-\alpha} & -2W_0^{'} & W_b^{'} \\
J_{-\alpha-\beta} & 0 & W_0^{'} \\
\end{array}
\right),
\]
the generators $\{J_{-\alpha},J_{-\alpha-\beta}\}$ commuting with the other
$W'$ generators.
Moreover, the $W'$ generators can be related to the
$W$ ones previously defined in eq. (\ref{7.1}):
\[
W_0^{'}=W_0 \qquad W_{a}^{'}=W_{a} \qquad
W_{b}^{'}=W_{b} +3 \frac{J_{-\alpha}}{J_{-\alpha-\beta}} W_{0} \qquad
W_{ab}^{'} = W_{ab} - \frac{J_{-\alpha}}{J_{-\alpha-\beta}} W_{a}
\]
Thus we recognize the $\cw_3^{(2)}$ algebra and
\[
Com(J_{-\alpha},J_{-\alpha-\beta})=\cw_3^{(2)} \oplus
\{J_{-\alpha},J_{-\alpha-\beta}\}.
\]

\newpage

\subsubsection{$\cw^{(1)}_3$ algebra}
We conclude this paragraph by recalling that the $\cw_3^{(1)}$ algebra
can itself be seen as the commutant of the whole $s \ell (3)$ algebra,
and even simply as
$Com(\hat{J}_1,\hat{J}_2,\hat{J}_{\alpha},\hat{J}_{-\alpha},\hat{J}_{-\beta},
\hat{J}_{-\alpha-\beta})$. So the $\cg_-^{s \ell (3)}$ part has been
enlarged with the help of three other generators (see section \ref{Tgle}).
We recall the two $\cw_3^{(1)}$ generators (\ie $s \ell (3)$ Casimirs):
\beano
\hat{C}_2 &=& s_2(\hat{J}_1,\hat{J}_1) +
\frac{3}{4}s_2(\hat{J}_2,\hat{J}_2) +
s_2(\hat{J}_{\alpha},\hat{J}_{-\alpha}) +
    s_2(\hat{J}_{\beta},\hat{J}_{-\beta}) +
s_2(\hat{J}_{\alpha+\beta},\hat{J}_{-\alpha-\beta})\\
\hat{C}_3 &=& \frac{1}{4}s_3(\hat{J}_2,\hat{J}_2,\hat{J}_2)
-s_3(\hat{J}_1,\hat{J}_1,\hat{J}_2)
-s_3(\hat{J}_{\alpha+\beta},\hat{J}_{-\alpha},\hat{J}_{-\beta})
-s_3(\hat{J}_{-\alpha-\beta},\hat{J}_{\alpha},\hat{J}_{\beta})+
\\
&& -s_3(\hat{J}_1,\hat{J}_{\alpha},\hat{J}_{-\alpha})
+s_3(\hat{J}_1,\hat{J}_{\beta},\hat{J}_{-\beta})
+\frac{1}{2}s_3(\hat{J}_2,\hat{J}_{\alpha},\hat{J}_{-\alpha})
+\frac{1}{2}s_3(\hat{J}_2,\hat{J}_{\beta},\hat{J}_{-\beta})+
\\
&& -s_3(\hat{J}_2,\hat{J}_{\alpha+\beta},\hat{J}_{-\alpha-\beta})
\enano
It is worthwhile to note that $\cw_3^{(1)}$ can also be obtained by
performing a symmetry fixing on the $\cw_3^{(2)}$ algebra in a way analogous
to the one used in $\cal G$ algebra (see section \ref{watts}). Such a technics
has been
used in \cite{Wgd} in the case of classical affine $\cal W$ algebras in
presence
of constraints. This method of secondary reduction had been also denoted
a gauging of a $\cal W$ algebra. A natural gradation shows up on the
C.R. (\ref{7.2}) of the $\cw_3^{(2)}$ algebra, with the zero graded part
generated by $\hat{Y}$ and $\hat{C}_2$. Performing first computations at
the classical level, one will finally obtain $\cw_3^{(1)}$ as the commutant
of the
$\cw_3^{(2)}$ subalgebra generated by $\hat{Y}$ and $\hat{W}_-$, \ie
\[
\cw_3^{(1)} = Com_{\cw_3^{(2)}}(\hat{Y},\hat{W}_-).
\]

\subsection{$s \ell (4)$ case}
Let $\alpha,\beta,\gamma$ be the simple positive roots.
\[
J = \left(
\begin{array}{cccc}
\frac{J_1+J_2}{2} & J_{\alpha} & J_{\alpha+\beta} & J_{\alpha+\beta+\gamma} \\
J_{-\alpha} & \frac{-J_1+J_2}{2} & J_{\beta} & J_{\beta+\gamma} \\
J_{-\alpha-\beta} & J_{-\beta} & \frac{-J_2+J_3}{2} & J_{\gamma} \\
J_{-\alpha-\beta-\gamma} & J_{-\beta-\gamma} & J_{-\gamma} &
\frac{-J_2-J_3}{2} \\
\end{array}
\right)
\]
In $s \ell (4)$, there are four different types of regular subalgebras
(up to conjugation):
\[
\ch = \su \, , \,  \su \oplus \su \, , \, s \ell (3) \, , \, s
\ell (4).
\]
 They correspond to four different $\su$-embeddings and so to
four different gradations,
$\cg=\cg_-^{\cal H} \oplus \cg_0^{\cal H} \oplus \cg_+^{\cal H}$. We denote the
corresponding $\cw(\cg,\ch)$ algebra as:
\[
\cw_4^{(1)} = \cw(s \ell (4),s \ell (4)) \qquad
\cw_4^{(2)} = \cw(s \ell (4),s \ell (3))
\]
\[
\cw_4^{(3)} = \cw(s \ell (4),\su \oplus \su) \qquad
\cw_4^{(4)} = \cw(s \ell (4),\su)
\]
In the following, we will present the quantum version, so dealing only
with $\hat{J}$ quantities. The Casimir of $s \ell (4)$ are denoted by
$\hat{C}_2$,$\hat{C}_3$,$\hat{C}_4$.

\subsubsection{$Com(\cg_-^{\cal H})$}
For each $\su$-embedding, we give:

\begin{itemize}
\item[-]
the corresponding regular subalgebra \ch,
\item[-]
the corresponding negative graded part of \cg: $\cg_-^{\cal H}$,
\item[-]
the number of steps needed for the symmetry fixing (see section
\ref{sect2}), and also the elements
$E_{-\varphi_i}$ of lowest grade,
\item[-]
an Abelian subalgebra of $\cg_-^{\cal H}$ with a maximal number of
generators: $\cg_-^-$
\item[-]
and finally the commutant of $\cg_-^{\cal H}$ identified as a
\cw-algebra plus the central part (after having performed a change of
basis).
\end{itemize}
\begin{center}
\begin{tabular}{||c|c|cc|c|c||}
\hline
\ch & $\cg_-^{\cal H}$ & steps & $E_{-\varphi_i}$ & $\cg_-^-$
& $Com(\cg_-^{\cal H})$ \\
\hline
$\su$ &
$\hat{J}_{-\alpha},\hat{J}_{-\alpha-\beta},\hat{J}_{-\alpha-\beta-\gamma}$
& 1 & $E_{-\alpha-\beta-\gamma}$
& $\hat{J}_{-\alpha},\hat{J}_{-\alpha-\beta}$
& $\cw_4^{(4)} \oplus \cg_-^-$ \\
 & & & & $\hat{J}_{-\alpha-\beta-\gamma}$ & \\
\hline
$2\ \su$ &
$\hat{J}_{-\beta},\hat{J}_{-\alpha-\beta},\hat{J}_{-\beta-\gamma}$
& 1 & $E_{-\alpha-\beta}+E_{-\beta-\gamma}$
& $\hat{J}_{-\beta},\hat{J}_{-\alpha-\beta}$ &
$\cw_4^{(3)} \oplus \cg_-^-$ \\
 & $\hat{J}_{-\alpha-\beta-\gamma}$ & &
 & $\hat{J}_{-\beta-\gamma},\hat{J}_{-\alpha-\beta-\gamma}$ & \\
\hline
$s \ell (3)$ & $\hat{J}_{-\alpha},\hat{J}_{-\gamma},\hat{J}_{-\alpha-\beta}$
& 1 & $E_{-\alpha-\beta-\gamma}$
& $\hat{J}_{-\alpha},\hat{J}_{-\alpha-\beta}$
& $\cw_4^{(4)} \oplus \{ \hat{J}_{-\alpha-\beta-\gamma} \} $ \\
 & $\hat{J}_{-\beta-\gamma},\hat{J}_{-\alpha-\beta-\gamma}$ & &
 & $\hat{J}_{-\alpha-\beta-\gamma}$ & \\
\hline
$s \ell (4)$ & $\hat{J}_{-\alpha},\hat{J}_{-\beta},\hat{J}_{-\gamma}$
& 2 & $E_{-\beta},E_{-\alpha-\beta-\gamma}$
& $\hat{J}_{-\beta},\hat{J}_{-\alpha-\beta}$
& $\cw_4^{(3)} \oplus $ \\
 &
$\hat{J}_{-\alpha-\beta},\hat{J}_{-\beta-\gamma},\hat{J}_{-\alpha-\beta-\gamma}$ & &
 & $\hat{J}_{-\beta-\gamma},\hat{J}_{-\alpha-\beta-\gamma}$
 & $\{ \hat{J}_{-\alpha-\beta-\gamma},C_w \}$ \\
\hline
\end{tabular}
\end{center}
where $C_w=\hat{J}_{-\beta}\hat{J}_{-\alpha-\beta-\gamma}
-\hat{J}_{-\alpha-\beta}\hat{J}_{-\beta-\gamma}$.
For $\ch =s \ell (4)$, $\cg_-^{\cal H}$ is the maximal nilpotent algebra in $s
\ell (4)$.
As expected from section \ref{sect2}, by considering the commutant of
the $\cg_-^{\cal H}$-part, we have obtained up to center only the algebra
$\cw_4^{(4)}$ for $\ch =\su,s \ell (3)$, and $\cw_4^{(3)}$ for $\ch=2 \su,s
\ell (4)$.
The difference between the case $\ch=\su$ and $\ch=s \ell (3)$
(respectively $\ch=2\su$ and $\ch=s \ell (4)$) stands in the center: the bigger
$\ch$,
the smaller the center.
\subsubsection{$Com(\cn)$ \label{nilpotent}}
Here we consider some other nilpotent algebras $\cal N$, not associated to an
$\su$ embedding.
\begin{itemize}
\item $\cn^{(I)}=\{
\hat{J}_{-\alpha},\hat{J}_{-\alpha-\beta-\gamma},\hat{J}_{-\gamma} \}$
\\
The commutant of $\cn^{(I)}$ is $\cw_4^{(4)} \oplus \cn^{(I)}$. This has
to be compared
with the commutant of $\cg_-^{s\ell(2)}$: $\cn^{(I)}$ and $\cg_-^{s\ell(2)}$
are
isomorphic nilpotent
(abelian) algebras, but they are not conjugated.
\item $\cn^{(II)}=\{ \hat{J}_{-\alpha},\hat{J}_{-\alpha-\beta},
\hat{J}_{-\beta-\gamma},\hat{J}_{-\alpha-\beta-\gamma} \}$
\\ $\cn^{(II)}$ has the same number of
generators as $\cg_-^{2s\ell(2)}$. However, its commutant is not
$\cw_4^{(3)}$ up to center, but
$\cw_4^{(4)} \oplus \{ \hat{J}_{-\alpha-\beta-\gamma} ,
\hat{J}_{-\alpha-\beta} \}$.
Note that in this case, an Abelian $\cal N$-subalgebra has at most
three generators, as in the cases $\cg_-^{s\ell(2)}$ and
$\cg_-^{s \ell (3)}$, leading to the same algebra $\cw_4^{(4)}$.
\item $\cn^{(III)}=\{
\hat{J}_{-\alpha},\hat{J}_{-\beta},\hat{J}_{-\alpha-\beta},
\hat{J}_{-\beta-\gamma},\hat{J}_{-\alpha-\beta-\gamma} \}$\\
$\cn^{(III)}$ has the same number of
generators as $\cg_-^{s \ell (3)}$, but the commutant of $\cn_-^{(III)}$
is $\cw_4^{(3)} \oplus \{
\hat{J}_{-\alpha-\beta},\hat{J}_{-\alpha-\beta-\gamma},C_w \}$.
As in the previous case, we note that the maximal abelian subalgebra in
$\cn^{(III)}$
is the same as in the cases $\cg_-^{2 s\ell(2)}$ and $\cg_-^{s \ell (4)}$,
where
their commutant lead also to the algebra $\cw_4^{(4)}$, up to center.
\end{itemize}
We remark that in each case, the commutant of a nilpotent algebra $\cal N$ has
a center formed by elements of $\cub(\cg_-^-)$, where $\cg_-^-$ is an Abelian
algebra contained in $\cal N$ with a maximal number of generators. Note
also that
the dimension of this center is $2 dim(\cn) - dim(\cg_-^-)$. We have seen that
algebras associated to different $\cal N$, but with same $\cg_-^-$-parts differ
from central elements. Thus, we can say that the role of $\cn \setminus
\cg_-^-$
is to select in $\cub(\cg_-^-)$ the elements that will form the center of
the commutant.

\subsubsection{$\cw_4^{(i)}$ as $Com(\tilde{\cg})$}
In the following, we describe all the $\cw(s \ell (4),\ch)$ algebras
in terms of the commutant of an $s \ell (4)$ subalgebra: note
that this subalgebra $\tilde{\cg}$ is not necessarily solvable. Following
section \ref{Tgle}, to compute this commutant,
we act first with the dual of $\tilde{\cg}_{\geq 0}$. All $J_{<0}^g$ are
therefore fixed: the
non-zero fixing conditions are presented in the column "fixing": they are
identical to the set of constraints used in the Hamiltonian reduction
framework. Then,
we act with the $\cg_+$-part in a way similar to the so-called highest
weight gauge. Finally, we enumerate the different generators, with their
degree.
\begin{center}
\begin{tabular}{||c|c|cc|cc||}
\hline
$\cw$ & $\tilde{\cg}$ & $\tilde{\cg}_{\geq 0}$ & fixing
& generators & degree \\
\hline
$\cw_4^{(4)}$ & $\hat{J}_{-\alpha},\hat{J}_{-\gamma}$
& $\hat{J}_{-\alpha},\hat{J}_{-\gamma}$
& $J_{4,1}^g=1$
& $\hat{H}_1,\hat{H}_2,\hat{E},\hat{F}$ & 1 \\
 & $\hat{J}_{-\alpha -\beta},\hat{J}_{-\beta -\gamma},\hat{J}_{-\alpha
-\beta -\gamma}$
 & $\hat{J}_1+\hat{J}_2+\hat{J}_3$ &
 & $\hat{G}_1^+,\hat{G}_1^-,\hat{G}_2^+,\hat{G}_2^-$ & $\frac{3}{2}$ \\
 & $\hat{J}_1+\hat{J}_2+\hat{J}_3$ & & & $\hat{C}_2$ & 2 \\
\hline
$\cw_4^{(3)}$ & $\hat{J}_{-\alpha},\hat{J}_{-\beta},\hat{J}_{-\gamma}$
& $\hat{J}_{-\alpha},\hat{J}_{-\gamma}$
& $J_{4,1}^g=1,J_{3,2}^g=1$
& $\hat{H},\hat{J}^+,\hat{J}^-$ & 1 \\
 & $\hat{J}_{-\alpha -\beta},\hat{J}_{-\beta -\gamma},\hat{J}_{-\alpha
-\beta -\gamma}$
 & $\hat{J}_2,\hat{J}_1+\hat{J}_3$
 & & $\hat{C}_2,\hat{S}_0,\hat{S}_+,\hat{S}_-$ & 2 \\
  & $\hat{J}_2,\hat{J}_1+\hat{J}_3$ & & & & \\
\hline
$\cw_4^{(2)}$ & $\hat{J}_{-\alpha},\hat{J}_{-\beta},\hat{J}_{-\gamma}$
& $\hat{J}_{-\beta},\hat{J}_{\alpha}$
& $J_{2,1}^g=1,J_{3,1}^g=1$
& $\hat{U}$ & 1 \\
 &
$\hat{J}_{-\alpha-\beta},\hat{J}_{-\beta-\gamma},\hat{J}_{-\alpha-\beta-\gamma}$
 & $\hat{J}_1,\hat{J}_2,\hat{J}_3$
 & $J_{4,2}^g=1$ & $\hat{C}_2,\hat{G}^+,\hat{G}^-$ & 2 \\
  & $\hat{J}_1,\hat{J}_2,\hat{J}_3,\hat{J}_{\alpha}$
  & & & $\hat{C}_3$ & 3 \\
\hline
$\cw_4^{(1)}$ & $\hat{J}_{-\alpha},\hat{J}_{-\beta},\hat{J}_{-\gamma}$
& $\hat{J}_{\alpha},\hat{J}_{\beta},\hat{J}_{\alpha+\beta}$
& $J_{2,1}^g=1$ & $\hat{C}_2$ & 2 \\
 &
$\hat{J}_{-\alpha-\beta},\hat{J}_{-\beta-\gamma},\hat{J}_{-\alpha-\beta-\gamma}$
 & $\hat{J}_1,\hat{J}_2,\hat{J}_3$
 & $J_{3,2}^g=1$ & $\hat{C}_3$ & 3 \\
  & $\hat{J}_1,\hat{J}_2,\hat{J}_3$
  & & $J_{4,3}^g=1$ & $\hat{C}_4$ & 4 \\
   & $\hat{J}_{\alpha},\hat{J}_{\beta},\hat{J}_{\alpha+\beta}$ & & & & \\
\hline
\end{tabular}
\end{center}
We reproduce hereafter the C.R. of these $s \ell (4)$ $\cw$ algebras
\cite{Tjin}and give
the explicit expression of their Casimir-like generators, using our
determinant trick.

\newpage
\begin{itemize}
\item $\cw_4^{(4)} \simeq s \ell (3)_{def} \oplus \hat{C}_2$.
\\
$\{ \hat{H}_1,\hat{E},\hat{F} \} \oplus \{ \hat{H}_2 \}$ forms a $\su \oplus
u(1)$ subalgebra.
$\{ \hat{G}_1^{\pm} \}$ and $\{ \hat{G}_2^{\pm} \}$ transform as two
$D_\frac{1}{2}$
representations under this $\su$, indices $1$ and $2$ refering to the
$u(1)$ charge. $\hat{C}_2$ is a central element.
\[\begin{array}{ll}
 [\hat{G}_1^+,\hat{G}_2^+]=-9\hat{H}_2\hat{E} \mb{}&
[\hat{G}_1^+,\hat{G}_2^-]=\frac{3}{16} \hat{C}_2 -\frac{9}{2} \hat{C}
-\frac{27}{4}\hat{H}_2^2 -9 \hat{H}_1\hat{H}_2
\\ & \\
{[\hat{G}_1^-,\hat{G}_2^-]}=9\hat{H}_2\hat{F} &
[\hat{G}_1^-,\hat{G}_2^+]=\frac{3}{16} \hat{C}_2 -\frac{9}{2} \hat{C}
-\frac{27}{4}\hat{H}_2^2
+9 \hat{H}_1\hat{H}_2
\end{array}
\]
where $\hat{C}=2\hat{H}_1^2+\hat{E}\hat{F}+\hat{F}\hat{E}$ is the Casimir of
the $\su$ subalgebra.
This \cw-algebra has two Casimir-like elements, which are proportional to
$\hat{C}_3$ and
$\hat{C}_4$, the third and fourth order Casimir of $s \ell (4)$:
\beano
\hat{C}_3 & \propto &  2s_2(\hat{G}_1^+,\hat{G}_2^-)
+2s_2(\hat{G}_1^-,\hat{G}_2^+) -9 \hat{H}_2 \hat{C}
+\frac{3}{8} \hat{H}_2 \hat{C}_2 -\frac{9}{2}\hat{H}_2^3 -9 \hat{H}_2 \\
\hat{C}_4 & \propto &  6s_3(\hat{H}_1,\hat{G}_1^+,\hat{G}_2^-)
+3s_3(\hat{H}_2,\hat{G}_1^+,\hat{G}_2^-)
+6s_3(\hat{H}_1,\hat{G}_1^-,\hat{G}_2^+)
+3s_3(\hat{H}_2,\hat{G}_1^-,\hat{G}_2^+) +\\
&&+6s_3(\hat{E},\hat{G}_1^-,\hat{G}_2^-) +6s_3(\hat{F},\hat{G}_1^+,\hat{G}_2^+)
-\frac{9}{16} \hat{C}_2 \hat{C} +\frac{9}{32} \hat{C}_2 \hat{H}_2^2
+\frac{54}{8} \hat{C}^2 +\frac{27}{4} \hat{C} \hat{H}_2^2+\\
&& -\frac{81}{16} \hat{H}_2^4+9 \hat{C} -\frac{27}{4} \hat{H}_2^2
\enano
\item $\cw_4^{(3)} \simeq (\su \oplus \su)_{def} \oplus \hat{C}_2$.
\\
$\{ \hat{H},\hat{J}^+,\hat{J}^- \}$ forms a $\su$ subalgebra. $\{
\hat{S}_0,\hat{S}_+,\hat{S}_- \}$ transforms
as a $D_1$ representation under the $\su$ subalgebra. $\hat{C}_2$ is a central
element.
\[
 [\hat{S}_0,\hat{S}_+]=(\hat{C}_2-\hat{C}')\hat{J}^+ \qquad
[\hat{S}_0,\hat{S}_-]=-(\hat{C}_2-\hat{C}')\hat{J}^-
\qquad [\hat{S}_+,\hat{S}_-]=2(\hat{C}_2-\hat{C}')\hat{H}
\]
where $\hat{C}'=2\hat{H}^2+\hat{J}^+\hat{J}^-+\hat{J}^-\hat{J}^+$ is the
Casimir of the $\su$ subalgebra.
\beano
\hat{C}_3 &\propto& \hat{H}\hat{S}_0 +\frac{1}{2}s_2(\hat{J}^-,\hat{S}_+)
+\frac{1}{2}s_2(\hat{J}^+,\hat{S}_-)
\\
\hat{C}_4 &\propto& 2s_2(\hat{S}_+,\hat{S}_-) +2 \hat{S}_0^2 +\hat{C}_2
\hat{C}' -\frac{1}{2}\hat{C}'^2 -2\hat{C}'
\enano
  \item $\cw_4^{(2)} \simeq \su_{def} \oplus \hat{C}_2 \oplus \hat{C}_3$.
\\
$\hat{C}_2$ and $\hat{C}_3$ are two central elements.
\[\begin{array}{l}
 [\hat{U},\hat{G}^{\pm}]=\pm \hat{G}^{\pm} \qquad
[\hat{G}^+,\hat{G}^-]=-\frac{1}{3} \hat{C}_3 -\hat{C}_2\hat{U} +\hat{U}^3
-\hat{U} \\ \\
\hat{C}_4 \propto 2s_2(\hat{G}^+,\hat{G}^-)+ \frac{1}{2}
\hat{U}^4-\hat{C}_2\hat{U}^2-\frac{2}{3}\hat{C}_3\hat{U}-\frac{1}{2}\hat{U}^2
\end{array}
\]
\item $\cw_4^{(1)} \simeq \hat{C}_2 \oplus \hat{C}_3 \oplus \hat{C}_4$.
\\
This is the well-known Casimir algebra of $s \ell (4)$: a commutative algebra.
\end{itemize}

\subsubsection{Commutant of second order}
It might be interesting to present, as already done in the previous paragraph
for $\cw_3^{(1)}$ and $\cw_3^{(2)}$, some relations between $\cw_4$
algebras using the above mentioned secondary reduction scheme. Their
above given C.R. allow to recognize natural gradings in each of these
algebras. Working first at the classical level, it will be possible to
realize some of them as the commutant of a part of another $\cw_4$ one.
In particular we have:
\[\begin{array}{l}
\cw_4^{(3)} = Com_{\cw_4^{(4)}}(\hat{H}_1,\hat{E})
\\
\cw_4^{(2)} = Com_{\cw_4^{(4)}}(\hat{H}_2,\hat{E},\hat{G}_1^-,\hat{G}_1^+)
\\
\cw_4^{(2)} = Com_{\cw_4^{(3)}}(\hat{H},\hat{J}^+)
\end{array}
\]
One might remark that the finite quantum algebras $\cw_4^{(3)}$ and
$\cw_4^{(4)}$ can
be related in this framework, while their classical affine analogues cannot
be linked by $\cal W$-gauge transformations (see \cite{Wgd}).

\hfill\break

\sect{Affine case \label{aff}}

\indent

The generalization of the above results when \cg\ is now an affine Kac-Moody
algebra is rather straightforward. Indeed recalling (classical case) the
transformation eq. (\ref{eq:4})
$$J \rightarrow J^g = g \ J g^{-1} + k \prt g . g^{-1}$$
one notes that the affine term $k \prt g . g^{-1}$ is in $\cg_+$ with
$g \in G_+$, which ensures that our symmetry fixing of sections \ref{Abel}
and \ref{sect2} directly applies. As developed in section \ref{sect2.3},
one thus obtains the algebras $\cw(\cg,\mu \su)$ up to center part. The
adjunction of Cartan elements to the $\cg_+$ transformations -see section
\ref{norm}- can again be performed without difficulties in order to get
rid of this center part. However to complete the set of \cw\ algebras,
one will use the secondary reduction technics already mentioned in section
\ref{watts}. Thus, as in the finite case, one obtains a \cw\ algebra
from the computation of commutants. For the quantum case, we will
have first to symmetrize and regularize (\ie normal order) the classical
expressions of the $W$ elements. Then we adjust, in a unique way, the scalar
$k$-functions which appear in the $W$ expressions, by requiring
again \cw\ to be the commutant of a special
subalgebra, either of \cg\ or, if a secondary "reduction" is necessary, of
another \cw\ one.

Let us illustrate our methods in the simplest cases.
All quantities depend in the variable z. We denote by $s_j$ the {\it
regularized} (\ie normal ordered) and symmetrized
product of $j$ fields. To recover from the quantum expressions of the
$W$ elements their classical counterpart, one has just, forgetting the
$s_j$'s, to replace each $k$-polynomial by its highest degree monomial.
Note also that, as expected, one gets for each \cw\ algebra at the quantum
level the same central charge as the one obtained by the BRS cohomology
method.

\begin{itemize}
\item  {\bf the Virasoro $\cw_2$ algebra}:

As usual we start with $\cg=\su^{(1)}$ with elements $\hat{J}_1(z)$ and
$\hat{J}_{\pm \alpha}(z)$. We remark that the commutant of
$\hat{J}_{-\alpha}$ contains $\hat{J}_{-\alpha}$ itself. To avoid this
last field and determine uniquely $T$, we rather determine the commutant
of $\hat{J}_{-\alpha}$ and $\hat{J}_1$.

\beano
T &=&\frac{1}{4 (k+2)\ \hat{J}_{-\alpha}^2} \,  \, \left\{ \, 4
s_4(\hat{J}_1,\hat{J}_1,\hat{J}_{-\alpha},\hat{J}_{-\alpha})+4
s_4(\hat{J}_{\alpha},\hat{J}_{-\alpha},\hat{J}_{-\alpha},\hat{J}_{-\alpha})
\right.+\\
&& +4 (k+4)\, \left[ s_3(\hat{J}_1,\prt \hat{J}_{-\alpha},\hat{J}_{-\alpha})
-s_3(\prt \hat{J}_1,\hat{J}_{-\alpha},\hat{J}_{-\alpha})\right]
-2(k^2+\frac{7}{2}k+\frac{8}{3})\, s_2(\prt^2
\hat{J}_{-\alpha},\hat{J}_{-\alpha}) +\\
&& \left.+3(k^2+4k+\frac{14}{3})\, s_2(\prt \hat{J}_{-\alpha},\prt
\hat{J}_{-\alpha}) \, \right\}
\enano
the central charge is equal to
\[
c(k) = 1 -6 \frac{(k+1)^2}{k+2}.
\]

We recognize the central charge $c(p,q) = 1-6 \ \frac{(p-q)^2}{pq}$ of
the Virasoro minimal models corresponding to $(p,q) = (k+2,1)$.

Let us add that by taking the classical limit we recover exactly the
usual Sugawara expression in the part of $T$ without derivatives.

\item {\bf the Bershadsky $\cw_3^{(2)}$ algebra}:

Now we turn our attention to the $\cg=s \ell (3)^{(1)}$ cases. In the
following, we keep the notation of section \ref{sl3}. The commutant of
the negatively graded part of $\cg^*$ with respect to the principal
gradation provides the classical $\cw_3^{(2)}$ algebra plus a central term,
which can be thrown way by enlarging $\cg_-^*$ with the help of a Cartan
generator (cf section \ref{sl3}). As for the Virasoro algebra, the
quantization reduces, after regularization and symmetrization, to a tuning
of the k-coefficients obtained by imposing the OPE's between the $W$
generators and the extended $\cg_-$ algebra, itself generated by
$\hat{J}_1,\hat{J}_{-\alpha},\hat{J}_{-\beta},\hat{J}_{-\alpha-\beta}$, to
vanish.

\beano
W_+ &=& \frac{1}{2 \hat{J}_{-\alpha-\beta}^{\frac{3}{2}}} \,  \, \left\{
\, 2s_3(\hat{J}_1,\hat{J}_{-\beta},\hat{J}_{-\alpha-\beta})
+3s_3(\hat{J}_2,\hat{J}_{-\beta},\hat{J}_{-\alpha-\beta})
+2s_3(\hat{J}_{-\alpha},\hat{J}_{-\beta},\hat{J}_{-\beta}) +\right.\\
&&
\left. +2s_3(\hat{J}_{\alpha},\hat{J}_{-\alpha-\beta},\hat{J}_{-\alpha-\beta})
+2(k+\frac{5}{2}) \left[ s_2(\hat{J}_{-\beta},\prt \hat{J}_{-\alpha-\beta})-
s_2(\prt \hat{J} _{-\beta},\hat{J}_{-\alpha-\beta})\right] \, \right\} \\
W_- &=& \frac{1}{2 \hat{J}_{-\alpha-\beta}^{\frac{3}{2}}} \,  \, \left\{
\, 2s_3(\hat{J}_1,\hat{J}_{-\alpha},\hat{J}_{-\alpha-\beta})
-3s_3(\hat{J}_2,\hat{J}_{-\alpha},\hat{J}_{-\alpha-\beta})
-2s_3(\hat{J}_{-\alpha},\hat{J}_{-\alpha},\hat{J}_{-\beta}) +\right.\\
&& \left.
+2s_3(\hat{J}_{\beta},\hat{J}_{-\alpha-\beta},\hat{J}_{-\alpha-\beta})
+2(k+\frac{5}{2}) \left[ s_2(\hat{J}_{-\alpha},\prt \hat{J}_{-\alpha-\beta})
-s_2(\prt \hat{J }_{-\alpha},\hat{J}_{-\alpha-\beta}) \right] \, \right\} \\
Y &=& \frac{1}{\hat{J}_{-\alpha-\beta}} \,  \,
\left\{
s_2(\hat{J}_2,\hat{J}_{-\alpha-\beta})+s_2(\hat{J}_{-\alpha},\hat{J}_{-\beta })
\, \right\} \\
T &=& \frac{1}{4(k+3) \hat{J}_{-\alpha-\beta}^2} \,  \, \left\{ \,
4s_4(\hat{J}_{\alpha+\beta},\hat{J}_{-\alpha-\beta},\hat{J}_{-\alpha-\beta}
,\hat{J}_{-\alpha-\beta})
+4s_4(\hat{J}_{\beta},\hat{J}_{-\beta},\hat{J}_{-\alpha-\beta},
\hat{J}_{-\alpha-\beta})
+\right.
\\
&& +4s_4(\hat{J}_{\alpha},\hat{J}_{-\alpha},\hat{J}_{-\alpha-\beta},
\hat{J}_{ -\alpha-\beta})
 +3s_4(\hat{J}_2,\hat{J}_2,\hat{J}_{-\alpha-\beta},\hat{J}_{-\alpha-\beta})
 +4s_4(\hat{J}_1,\hat{J}_1,\hat{J}_{-\alpha-\beta},\hat{J}_{-\alpha-\beta})
+\\
&&
+(k+5)\left[4s_3(\hat{J}_1,\hat{J}_{-\alpha-\beta},\prt
\hat{J}_{-\alpha-\beta})
-4s _3(\prt \hat{J}_1,\hat{J}_{-\alpha-\beta},\hat{J}_{-\alpha-\beta})
+2s_3(\hat{J}_{-\alpha},\prt \hat{J}_{-\beta},\hat{J}_{-\alpha-\beta})+
\right.\\
&& \left. +2s_3(\prt
\hat{J}_{-\alpha},\hat{J}_{-\beta},\hat{J}_{-\alpha-\beta}) \right]
+3(k^2+\frac{31}{6}k+\frac{43}{6})s_2(\prt \hat{J}_{-\alpha-\beta},\prt
\hat{J}_{-\alpha-\beta})+
\\
&&\left. -2(k^2+\frac{9}{2}k+\frac{23}{6})\, s_2(\hat{J}_{-\alpha-\beta},
\prt^2 \hat{J}_{-\alpha-\beta}) \, \right\}
\enano
the central charge is equal to
\[
c(k) = 1 -6 \frac{(k+1)^2}{k+3}.
\]
which correspond to the series $c(p,q)=1-12\frac{(p-q)^2}{pq}$ presented in
\cite{Bersh} for
$(p,q)=(k+3,2)$.

\item {\bf the Zamolodchikov $\cw_3^{(1)}$ algebra}:

The classical $\cw_3^{(1)}$ algebra can be obtained by a secondary
"reduction" \cite{Wgd} of the $\cw_3^{(2)}$ algebra, or in other words by
determining
the commutant in the (closure of the) enveloping algebra of $\cw_3^{(2)}$
of the $W_-$ and $Y$ generators. Once again, quantization is achieved,
after introduction of the $s_j$ products, by imposing the OPE's between
the $\cw_3^{(1)}$ generators and the previous $W_-$ and $Y$ ones to be
zero, determining in this way the $k$-dependent quantum corrections.
Hereafter, $W_2$ and $W_3$ are written as functions of the $\cw_3^{(2)}$
generators. Their expressions in terms of $s \ell (3)^{(1)}$ generators
are obviously obtained by using the above given $\cw_3^{(2)}$ realization.

\beano
W_3 &=& \sqrt{\frac{-3}{8(3k+4)(5k+12)}} \frac{1}{(k+3) W_-^3} \, \, \left\{
\rule{0mm}{5mm} \, s_5(W_+,W_-,W_-,W_-,W_-)
+4s_6(Y,Y,Y,W_-,W_-,W_-)+ \right.\\
&&  -4(k+3)s_5(T,Y,W_-,W_-,W_-)
-6(k+9)\, s_5(\prt Y,Y,W_-,W_-,W_-)+ \\
&& +4(k+\frac{15}{2})\, s_5(Y,Y,\prt W_-,W_-,W_-)
 +2(k+6)(k+3)\, s_4(\prt T,W_-,W_-,W_-)+ \\
&& -(\frac{8k}{3}+14)(k+3)\, s_4(T,\prt W_-,W_-,W_-)
+(k^2+2k+25)\, s_4(\prt^2 Y,W_-,W_-,W_-)+ \\
&& -(2k^2+15k+33)\, s_4(\prt Y,\prt W_-,W_-,W_-)
-(2k+5)(k+3)\, s_4(Y,\prt^2 W_-,W_-,W_-)+ \\
&& +2(\frac{5}{3}k^2+13k+33)\, s_4(Y,\prt W_-,\prt W_-,W_-)
+(\frac{2}{3}k^3+5k^2+\frac{103}{9}k+\frac{7}{3})\, s_3(\prt^3 W_-,W_-,W_-)+ \\
&& -\frac{1}{3}(10k^3+76k^2+181k+129)\, s_3(\prt^2 W_-,\prt W_-,W_-)+ \\
&& \left.+\frac{1}{3}(\frac{80}{9}k^3+73k^2+205k+207)\, s_3(\prt W_-,\prt
W_-,\prt W_-)
\, \right\} \, \\
W_2 &=& \frac{1}{(k+3) W_-^2} \, \, \left\{ \,(k+3) s_3(T,W_-,W_-)
+(k+3)(-\frac{3}{2}s_3(\prt Y,W_-,W_-)+s_3(Y,\prt W_-,W_-))+ \right.\\
&&\left.+(\frac{4}{3}k^2+\frac{15}{2}k+\frac{21}{2})s_2(\prt W_-,\prt W_-)
-(k^2+\frac{23}{4}k+\frac{33}{4})s_2(\prt^2 W_-,W_-) \, \right\} \,
\enano

the central charge is equal to
\[
c(k) = 2(1-12 \frac{(k+2)^2}{k+3})
\]

and we recognize the $\cw_3$ minimal models
\[
c(p,q) = 2 \left( 1-12 \frac{(p-q)^2}{pq} \right) \mb{with} (p,q)
= (k+3,1)
\]

Again, it is straightforward to obtain in the classical limit the
Casimir expressions of $W_2$ and $W_3$ when restricting their expressions
to terms without derivatives.
\end{itemize}

\hfill\break

\sect{Conclusion}
We have shown that each \cw\ algebra, which arises from the usual
Hamiltonian reduction, can explicitely be realized in terms of all the $J$
components in the classical and quantum cases. The primary $W$ fields
are no longer polynomials, but quotients of two polynomials. However the
denominator quantities simply commute with all the numerator ones, allowing
in particular to compute OPE without special difficulties in the quantum
framework.

\indent
Actually, such an approach might be seen as a kind of generalized Sugawara
construction. Considering, as an example, the Zamolodchikov $\cw_3^{(1)}$
algebra realization given in section \ref{aff}, it seems natural to
compare it with the Casimir algebra approach of \cite{B1}. We
remark that in our case, we are not restricted to the level value $k=1$
of the $s \ell (3)^{(1)}$ algebra in order to avoid the occurence of extra
fields and need not the help of the coset \cite{B2}
$(s \ell (3)^{(1)}_k \oplus s \ell (3)^{(1)}_1 )/ s \ell (3)^{(1)}_{k+1}$.
Moreover, let us emphasize that the developed technics in this
paper do not limit to the Casimir-like $\cw_n$ algebras, but to any
$\cw(\cg,\ch)$ algebra following the notation used in our text: see again in
section \ref{aff} the realization of the Bershadsky $\cw_3^{(2)}$ algebra.

\indent
Let us also note that the generalization of such technics to (affine)
superalgebras \cite{F} is a priori straightforward. Finally, it might be
worthwhile to remind that in our construction the \cw\ algebra appears as
the commutant of a subalgebra. In this aspect, one can think in terms of
a coset construction. Such a mathematical object, which is explicitely
determined,
can be of interest by itself. For example, in the finite case, it has
been shown that $Com(\cg_-)$ with $\cg_-$ Abelian allows to construct
general realizations of the algebra \cg\ once a special \cg\
representation is known-cf \cite{BRS}.

\newpage

\appendix
\sect{Appendix: Symmetry fixing for the principal gradation of $so(n)$ and
$sp(2n)$ algebras}

\indent

Before considering these different algebras, let us come back to the \sln\
case developed in section
\ref{Abel}. We denote by $E_{-\psi_0}$ the smallest root generator, with
$\psi_0= \alpha_1+\alpha_2+\dots+\alpha_{n-1}$, \ie the sum of simple roots
in \sln. From now
on, we will work on the $n\times n$ matrix representation of \sln.
Then, the generator $E_{-\psi_0}$  is represented by the matrix $e_{n,1}$.

At the first
step, the positive root generators acting on $E_{-\psi_0}$ and which
constitue $\cg_+^{(0)}$ (see
eq. \ref{g+0}) are those with non-zero entries on the first row or on the last
column of the $n\times
n$ matrix, except the two end points on the
diagonal. One can
therefore put to zero all the entries of the first column and the last row,
included the Cartan generator
$H_{\psi_0}=e_{1,1}-e_{n,n}$ on the diagonal, and
excepted $E_{-\psi_0}=e_{n,1}$ itself.

At the second step, one acts on $E_{-\psi_1}$ with $\psi_1=
\alpha_2+\dots+\alpha_{n-2}$, and then describes the square
$(n-2)\times(n-2)$ with the point
$(n-1,2)$ as the low-left corner. Iterating the process, one is led to the
nested boxes representented
in figure \ref{carre}. Note that the general element $E_{-\psi_j}$, $0\leq
j\leq [\frac{n}{2}]$
corresponds to the root
\be
\begin{array}{l}
\psi_j= \alpha_{j+1}+ \alpha_{j+2}+ \dots+ \alpha_{n-j-1} \mb{for} 0\leq j\leq
\left[\frac{n}{2}\right]-1 \\
\psi_{[\frac{n}{2}]}=
\left\{\begin{array}{l} \alpha_{\frac{n}{2}} \mb{if} n\mbox{ even}\\
\alpha_{\frac{n-1}{2}}+\alpha_{\frac{n+1}{2}} \mb{if} n\mbox{
odd}\end{array}\right.
\end{array}
\ee
These elements are the only ones below the diagonal that have not been set
to zero in the
process of fixing the symmetry. On the diagonal, we have set to zero all
the Cartan generators
$H_{\psi_j}=e_{j,j}-e_{n+1-j,n+1-j}$ with $j=1,\dots,\mu=[\frac{n}{2}]$.
Thus, it remains
$n-1-\mu$ generators, which can be taken for instance of the form
$Y_j=e_{1,1}+e_{n,n}-e_{j,j}-e_{n+1-j,n+1-j}$ with $j=2,\dots,
[\frac{n+1}{2}]$.

\subsection{$so(2n+1)$ case}

\indent

It might be useful to remind that the $so(2n+1)$ algebra can be obtained by
a folding of the
$s\ell(2n+1)$ one. Indeed, considering the $(2n+1)\times(2n+1)$ matrix
representation of
$s\ell(2n+1)$:
\be
M= m^{ij} e_{ij} \mb{with} m^{ij}\in\R \mb{and} \sum_{i=1}^{2n+1}\ m^{ii}=0,
\label{sl2n+1}
\ee
the $so(2n+1)$ elements are $s\ell(2n+1)$ elements satisfying the condition
\be
m^{ij}= (-)^{i+j+1}\ m^{2n+2-j,2n+2-i}\ \ \forall \ i,j \label{so2n+1}
\ee
which reflects a (graded) symmetry w.r.t. the anti-diagonal.

Now, the smallest root generator reads
\be
E_{-\psi_0}= e_{2n,1}+e_{2n+1,2} \mb{with} \psi_0=\alpha_1+2\alpha_2+\dots
+2\alpha_n
\ee
The $\cg_+^{(0)}$ subalgebra appears to be made of elements with non-zero
entries on the two first
rows and two last columns, but without the element
$E_{\alpha_1}=e_{12}+e_{2n,2n+1}$ and the
diagonal. The corresponding parameters of symmetry transformations will be
fixed by the condition
$J|_{\cg^{(0)}_{\leq0}}=0$ where $\cg^{(0)}_{\leq0}$ is formed by the
elements with
with non-zero entries on the two first
rows and two last columns, included the diagonal (Cartan generator
$H_{\psi_0}=e_{1,1}-e_{2n+1,2n+1}$),
but without the element $E_{-\alpha_1}=e_{21}+e_{2n+1,2n}$ and the root
generator $E_{-\psi_0}$.

\begin{figure}[hbtp]
\begin{center}
\begin{picture}(204,204)
\put(0,0){\framebox(200,200)}
\put(20,0){\line(0,1){200}}
\put(0,20){\line(1,0){200}}
\put(180,0){\line(0,1){200}}
\put(0,180){\line(1,0){200}}

\put(20,0){\makebox(0,0){$\bullet$}}
\put(0,20){\makebox(0,0){$\bullet$}}
\put(180,0){\makebox(0,0){$\bullet$}}
\put(0,180){\makebox(0,0){$\bullet$}}
\put(0,20){\line(1,-1){20}}

\put(171,-9){\dashbox(38,38)}
\put(-9,171){\dashbox(38,38)}
\multiput(-10,169)(2,-2){90}{\tiny.}

\put(40,40){\framebox(120,120)}
\put(60,40){\line(0,1){120}}
\put(40,60){\line(1,0){120}}
\put(140,40){\line(0,1){120}}
\put(40,140){\line(1,0){120}}

\put(40,60){\makebox(0,0){$\bullet$}}
\put(60,40){\makebox(0,0){$\bullet$}}
\put(140,40){\makebox(0,0){$\bullet$}}
\put(40,140){\makebox(0,0){$\bullet$}}
\put(40,60){\line(1,-1){20}}

\put(135,35){\dashbox(30,30)}
\put(35,135){\dashbox(30,30)}
\multiput(34,133)(2,-2){50}{\tiny.}

\put(75,75){\framebox(50,50)}

\put(240,130){\vector(-1,0){40}}
\put(280,130){\makebox(0,0){$so(2n+1)$}}
\put(240,110){\vector(-1,0){80}}
\put(280,110){\makebox(0,0){$so(2n-3)$}}
\put(240,90){\vector(-1,0){115}}
\put(280,90){\makebox(0,0){$so(2n-7)$}}
\put(280,160){\makebox(0,0){$\bullet=E_{-\varphi_i}$}}

\end{picture}
\end{center}
\caption{Remaining non-zero entries on and below the diagonal, after fixing
of the
symmetry\label{paul}} \end{figure}
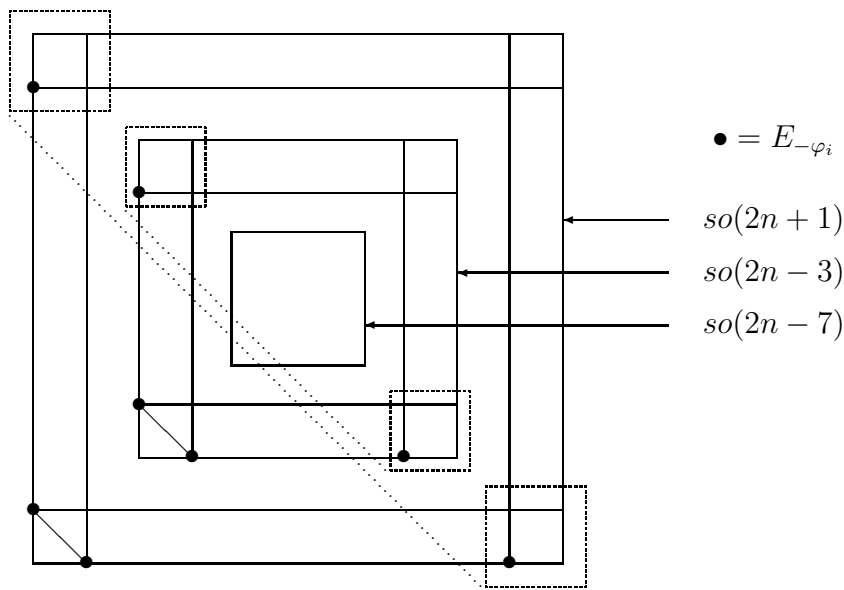

Then, one will have to consider, at the second step, the $so(2n-3)$
subalgebra made with
the $(2n-3)\times(2n-3)$ submatrix of low-left corner $(2n-1,3)$. One will
consider separately (in
the last step)
the $\su$ subalgebra, which commutes with $so(2n-3)$ and is constructed from
$E_{\alpha_1}$,
with negative root generator
$E_{-\alpha_1}= e_{2,1}+e_{2n+1,2n}$.
In the $so(2n-3)$ subalgebra, the lowest root generator is
\be
E_{-\psi_1}= e_{2n-2,3}+e_{2n-1,4} \mb{with}
\psi_1=\alpha_3+2\alpha_4+\dots +2\alpha_n
\ee
Iterating the process leads to the chain of embeddings $so(2n+1)\supset
so(2n-3)\oplus \su$,
$so(2n-3)\supset so(2n-7)\oplus \su$, and so on: see figure \ref{paul}. At
the last step, we will
deal with a direct sum of $\su$ subalgebras, and the parameters of the
corresponding root
generators $E_{\alpha_1}$, $E_{\alpha_3}$, $\dots$ will be fixed by
annihilating the
associated Cartan generators $H_{\alpha_1}$, $H_{\alpha_3}$, $\dots$

It follows that the
generators of non-positive grade which cannot be eliminated in the process
are the elements
$E_{-\psi_j}$ with $\psi_j$ in the set
\be
\begin{array}{l}
\alpha_1+2\alpha_2+\dots +2\alpha_n\mb{;}
\alpha_1 \mb{;} \alpha_3+2\alpha_4+\dots +2\alpha_n\\
\alpha_3 \mb{;} \alpha_5+2\alpha_6+\dots +2\alpha_n\mb{;}
\dots \mb{;}\left\{\begin{array}{ll}
\mbox{If $n$ odd } & \alpha_{n-2}\ ;\ \alpha_{n}\\
\mbox{If $n$ even } & \alpha_{n-1}\end{array}\right.
\end{array}
\ee

\subsection{$so(2n)$ case}

\indent

Deleting the $(n+1)^{th}$ row and the $(n+1)^{th}$ column in the $so(2n+1)$
matrix leads to a matrix realization of the $so(2n)$ algebra.
It is a simple exercice
to check that the same kind of nesting can be obtained, \ie $so(2n)\supset
so(2n-4)\oplus \su$,
$so(2n-4)\supset so(2n-8)\oplus \su$, and so on. The list of roots
corresponding to the remaining
(negatively graded) generators is now:
\[
\begin{array}{l}
\alpha_1+2\alpha_2+\dots +2\alpha_{n-2}+ \alpha_{n-1}+ \alpha_n \mb{;}
\alpha_1 \mb{;} \alpha_3+2\alpha_4+\dots +2\alpha_{n-2}+ \alpha_{n-1}+
\alpha_n\\
\alpha_3 \, ;\, \alpha_5+2\alpha_6+\dots +2\alpha_{n-2}+ \alpha_{n-1}+
\alpha_n\, ;\,
\dots \, ;\, \left\{\!\begin{array}{ll}
\alpha_{n-4};\, \alpha_{n-2}+ \alpha_{n-1}+ \alpha_n ;\,
\alpha_{n-2} ;\, \alpha_{n}&\mbox{if $n$ odd}\\
\alpha_{n-3} ;\, \alpha_{n-1}+ \alpha_n ;\,
\alpha_{n-2} ;\, \alpha_n & \mbox{if $n$ even}
\end{array}\right.
\end{array}
\]
We get the same picture as in figure \ref{paul} for the non-zero generators of
non-positive grade that
remain after fixing of the symmetry.

\subsection{$sp(2n)$ case}

\indent

The $sp(2n)$ algebra can be obtained from the folding of the $s\ell(2n)$
algebra. Again, we will
impose to the $s\ell(2n)$ elements a graded symmetry as we did above with
$s\ell(2n+1)$ elements in the $so(2n+1)$ case.
Then, as in the \sln\ case, the $\cg_+^{(0)}$ subalgebra will
be constituted by the first row and last
column with exception of the diagonal. We will have a situation analogous
to the one of figure
\ref{carre} for \sln, with the embeddings $sp(2n)\supset sp(2n-2)\supset
sp(2n-4)\supset\dots$.
Now, the set of roots corresponding to the negatively graded remaining
generators stand as follows:
\be
\begin{array}{l}
2\alpha_1+2\alpha_2+\dots +2\alpha_{n-1}+ \alpha_n \mb{;}
2\alpha_2+\dots +2\alpha_{n-1}+ \alpha_n\\
2\alpha_3+\dots +2\alpha_{n-1}+ \alpha_n\mb{;}
\dots \mb{;}2\alpha_{n-1}+ \alpha_n\mb{;}
\alpha_{n}
\end{array}
\ee

\end{document}